\documentclass[12pt,twoside]{article}
\usepackage{amssymb}
\usepackage{amsmath}
\usepackage{mathrsfs} 
\usepackage{amsfonts}
\usepackage[dvips]{graphicx}
\begin{document}
\newtheorem{d1}{Definition}[section]
\newtheorem{c1}{Corollary}[section]
\newtheorem{l1}{Lemma}[section]
\newtheorem{r1}{Remark}[section]

\newcommand{\cA}{{\cal A}}
\newcommand{\cB}{{\cal B}}
\newcommand{\cC}{{\cal C}}
\newcommand{\cD}{{\cal D}}
\newcommand{\cE}{{\cal E}}
\newcommand{\cF}{{\cal F}}
\newcommand{\cG}{{\cal G}}
\newcommand{\cH}{{\cal H}}
\newcommand{\cI}{{\cal I}}
\newcommand{\cJ}{{\cal J}}
\newcommand{\cK}{{\cal K}}
\newcommand{\cL}{{\cal L}}
\newcommand{\cM}{{\cal M}}
\newcommand{\cN}{{\cal N}}
\newcommand{\cO}{{\cal O}}
\newcommand{\cP}{{\cal P}}
\newcommand{\cQ}{{\cal Q}}
\newcommand{\cR}{{\cal R}}
\newcommand{\cS}{{\cal S}}
\newcommand{\cT}{{\cal T}}
\newcommand{\cU}{{\cal U}}
\newcommand{\cV}{{\cal V}}
\newcommand{\cX}{{\cal X}}
\newcommand{\cW}{{\cal W}}
\newcommand{\cY}{{\cal Y}}
\newcommand{\cZ}{{\cal Z}}

\def\bd{\begin{description}}
\def\be{\begin{enumerate}}
\def\ed{\end{description}}
\def\ee{\end{enumerate}} 
\def\al{\alpha}
\def\b{\beta}
\def\benrr{\begin{eqnarray*}}
\def\eenrr{\end{eqnarray*}}
\def\bc{\begin{center}}
\def\ec{\end{center}}
\def\d{\dot}
\def\si{\sigma}

\def\fr{\frac}
\def\sq{\sqrt}

\def\et{\end{tabular}}
\def\lf{\leftarrow}
\def\nn{\nonumber}
\def\va{\vartheta}
\def\wh{\widehat}
\def\vs{\vspace}
\def\la{\lambda}
\def\Lam{\Lambda} 
\def\sm{\setminus}
\def\ba{\begin{array}}
\def\ea{\end{array}} 
\def\bd{\begin{description}}
\def\ed{\end{description}}
\def\lan{\langle}
\def\ran{\rangle}

\bc

\textbf{ An Experimentally accessible geometric measure for entanglement in $N$-qubit pure states}

\vspace{.2 in} 

 \textbf{ Ali Saif M. Hassan\footnote{Electronic address: alisaif@physics.unipune.ernet.in} and  Pramod S. Joag\footnote{Electronic address: pramod@physics.unipune.ernet.in}}\\
 Department of Physics, University of Pune, Pune-411007, India.
 \ec

\vspace{.2in}

We present a multipartite entanglement measure for $N$-qubit pure states, using the norm of the correlation tensor which occurs in the Bloch representation of the state. We compute this measure for  several important classes of $N$-qubit pure states such as GHZ states, W states and their superpositions. We compute this measure for interesting applications like one dimensional Heisenberg antiferromagnet.  We use this measure to follow the entanglement dynamics of Grover's algorithm. We prove that this measure possesses almost all the properties expected of a good entanglement measure, including monotonicity. Finally, we extend this measure to $N$-qubit mixed states via convex roof construction  and establish its various properties, including its monotonicity. We also introduce a related measure which has all properties of the above measure and is also additive.\\


\vspace{.2in}
\bc
\textbf{I. INTRODUCTION}\\
\ec
\vspace{.2in}

Entanglement has proved to be a vital physical resource for various kinds of quantum information processing, including quantum state teleportation [1,2], cryptographic key distribution [3], classical communication over quantum channels [4,5,6], quantum error correction [7], quantum computational speedups [8] and distributed computation [9,10].
Further, entanglement is expected to play a crucial role in the many particle phenomena such as quantum phase transitions, transfer of information across a spin chain   [11,12] etc. Therefore, quantification of entanglement of multipartite quantum  states is fundamental to the whole field of quantum information and in general, to the physics of multicomponent quantum systems. Whereas the entanglement in pure bipartite states is well understood, classification of multipartite pure states and mixed states, according to the degree and character of their entanglement is still a matter of intense research [13,14.15]. Principal achievements are in the setting of bipartite systems. Among these, one highlights Wootter's formula for the entanglement of formation of two qubit mixed states [16], which still awaits a viable generalization to multiqubit case. Others include corresponding results for highly symmetric states [17,18,19]. The issue of entanglement in multipartite states is far more complex. Notable achievements in this area include applications of the relative entropy [20], negativity [21] Schimidt measure [22] and the global entanglement measure proposed by Meyer and Wallach [23].  

A measure of entanglement is a function on the space of states of a multipartite system, which is invariant on individual parts. Thus a complete characterization of entanglement is the characterization of all such functions. Under the most general local operations assisted by classical communication (LOCC), entanglement is expected to decrease. A measure of entanglement which decreases under LOCC is called an entanglement monotone. On bipartite pure states the sums of the $k$ smallest eigenvalues of the reduced density matrix are entanglement monotones. However, the number of independent invariants (i.e. the entanglement measures) increase exponentially as the number of particles $N$ increases and complete characterization rapidly becomes impractical. A pragmatic approach would be to seek a measure which is defined for any number of particles (scalable), which is easily calculated and which provides physically relevant information or equivalently, which passes the tests expected of a {\it good} entanglement measure [13,14].

In this paper, we present a global entanglement measure for $N$-qubit pure states which is scalable, which passes most of the tests expected of a good measure and whose value for a given system can be determined experimentally, without having a detailed {\it prior} knowledge of the state of the system. The measure is based on the Bloch representation of multipartite quantum states [24].

The paper is organized as follows. In section II we give the Bloch representation of a $N$-qubit quantum state and define our measure $E_{\mathcal{T}}.$ In section III we compute $E_{\mathcal{T}}$ for different classes of $N$-qubit states, namely, $GHZ$ and $W$ states and their superpositions. In section IV we prove various properties of $E_{\mathcal{T}},$ including its monotonicity, expected of a good entanglement measure. In section V we extend $E_{\mathcal{T}}$ to $N$-qubit mixed states via convex roof and establish its monotonicity. Finally, we conclude in section VI.

\bc

\textbf{II. BLOCK REPRESENTATION OF A $N$-QUBIT STATE AND THE DEFINITION OF THE MEASURE}\\
\ec

Consider the generators $\{I,\si_x,\;\si_y,\;\si_z\}\equiv \{\si_0,\si_1,\si_2,\si_3\}$ of $SU(2)$ group (Pauli matrices). These hermitian operators form a orthogonal basis (under the Hilbert-Schmidt scalar product) of the Hilbert space of operators acting on a single qubit state space. The $N$ times tensor product of this basis with itself generates a product basis of the Hilbert space of operators acting on the $N$-qubit state space. Any $N$-qubit density operator $\rho$ can be expanded in this basis. The corresponding expansion is called the Bloch representation of $\rho$ [24].

  In order to give the Bloch representation of a density operator acting on the Hilbert   
  space $\mathbb{C}^{2} \otimes \mathbb{C}^{2} \otimes \cdots \otimes \mathbb{C}^{2}$
  of a $N$-qubit quantum system, we introduce following notation. We use $k$, $k_i \; (i=1,2,\cdots)$ to denote a qubit chosen from $N$ qubits, so that $k$,\; $k_i \; (i=1,2,\cdots)$ take values in the set  $\mathcal{N}=\{1,2,\cdots,N\}$. The variables $\alpha_k \;\mbox{or} \; \alpha_{k_i}$ for a given $k$ or $k_i$ span the set of generators of $SU(2)$ group for the $k$th or $k_i$th qubit, namely the set $\{I_{k_i},\si_{1_{k_i}},\si_{2_{k_i}},\si_{3_{k_i}}\}$ for the $k_i$th qubit. For two qubit $k_1$ and $k_2$ we define
  
   $$\si^{(k_1)}_{\alpha_{k_1}}=(I_{2}\otimes I_{2}\otimes \dots \otimes \si_{\alpha_{k_1}}\otimes I_{2}\otimes \dots \otimes I_{2})   $$
   $$\si^{(k_2)}_{\alpha_{k_2}}=(I_{2}\otimes I_{2}\otimes \dots \otimes \si_{\alpha_{k_2}}\otimes I_{2}\otimes \dots \otimes I_{2})  $$
   $$\si^{(k_1)}_{\alpha_{k_1}} \si^{(k_2)}_{\alpha_{k_2}}=(I_{2}\otimes I_{2}\otimes \dots \otimes \si_{\alpha_{k_1}}\otimes I_{2}\otimes \dots \otimes \si_{\alpha_{k_2}}\otimes I_{2}\otimes I_{2})   \eqno{(1)}$$
   
  where  $\si_{\alpha_{k_1}}$ and $\si_{\alpha_{k_2}}$ occur at the $k_1$th and $k_2$th places (corresponding to $k_1$th and $k_2$th qubits respectively) in the tensor product and are the $\alpha_{k_1}$th and  $\alpha_{k_2}$th generators of $SU(2),\; (\alpha_{k_1}=1,2,3\; \mbox{and} \; \alpha_{k_2}=1,2,3)$ respectively. Then we can write

$$\rho=\fr{1}{2^N} \{\otimes_k^N I_{2}+ \sum_{\{k \}\subset \mathcal{N}}\sum_{\alpha_{k}}s_{\alpha_{k}}\si^{(k)}_{\alpha_{k}} +\sum_{\{k_1,k_2\}}\sum_{\alpha_{k_1}\alpha_{k_2}}t_{\alpha_{k_1}\alpha_{k_2}}\si^{(k_1)}_{\alpha_{k_1}} \si^{(k_2)}_{\alpha_{k_2}}+\cdots +$$
$$\sum_{\{k_1,k_2,\cdots,k_M\}}\sum_{\alpha_{k_1}\alpha_{k_2}\cdots \alpha_{k_M}}t_{\alpha_{k_1}\alpha_{k_2}\cdots \alpha_{k_M}}\si^{(k_1)}_{\alpha_{k_1}} \si^{(k_2)}_{\alpha_{k_2}}\cdots \si^{(k_M)}_{\alpha_{k_M}}+ \cdots+\sum_{\alpha_{1}\alpha_{2}\cdots \alpha_{N}}t_{\alpha_{1}\alpha_{2}\cdots \alpha_{N}}\si^{(1)}_{\alpha_{1}} \si^{(2)}_{\alpha_{2}}\cdots \si^{(N)}_{\alpha_{N}}\}.\eqno{(2)}$$

 where $\textbf{s}^{(k)}$ is a Bloch vector (see below) corresponding to $k$th subsystem,  $\textbf{s}^{(k)} =[s_{\alpha_{k}}]_{\alpha_{k}=1}^{3} $ which is a tensor of order one defined by
 $$s_{\alpha_{k}}= Tr[\rho \si^{(k)}_{\alpha_{k}}]= Tr[\rho_k \si_{\alpha_{k}}],\eqno{(3)}$$ where $\rho_k$ is the reduced density matrix for the $k$th qubit. Here $\{k_1,k_2,\cdots,k_M\},\; 1 \le M \le N,$ is a subset of $\mathcal{N}$ and can be chosen in $\binom{N}{M}$  ways, contributing $\binom{N}{M}$ terms in the sum $\sum_{\{k_1,k_2,\cdots,k_M\}}$ in Eq.(2), each containing a tensor of order $M$. The total number of terms in the Bloch representation of $\rho$ is $2^N$. We denote the tensors occurring in the sum $\sum_{\{k_1,k_2,\cdots,k_M\}},\; (1 \le M \le N)$ by $\mathcal{T}^{\{k_1,k_2,\cdots,k_M\}}=[t_{\alpha_{k_1}\alpha_{k_2}\cdots \alpha_{k_M}}]$ which  are defined by 
 
 $$t_{\alpha_{k_1}\alpha_{k_2}\dots\alpha_{k_M}}= Tr[\rho \si^{(k_1)}_{\alpha_{k_1}} \si^{(k_2)}_{\alpha_{k_2}}\cdots \si^{(k_M)}_{\alpha_{k_M}}]$$
 
$$ = Tr[\rho_{k_1k_2\dots k_M} (\si_{\alpha_{k_1}}\otimes\si_{\alpha_{k_2}}\otimes\dots \otimes\si_{\alpha_{k_M}})]   \eqno{(4)}$$

where $\rho_{k_1k_2\dots k_M}$ is the reduced density matrix for the subsystem $\{k_1 k_2\dots k_M\}$. We call  The tensor in last term in Eq. (2) $\mathcal{T}^{(N)}$.

From Eq.(4) we see that all the correlations between $M$ out of $N$ qubits are contained in  $\mathcal{T}^{\{k_1,k_2,\cdots,k_M\}}$ and all the $N$ qubit correlations are contained in  $\mathcal{T}^{(N)}$.

If $\rho$ is a $N$-qubit pure state we have $$Tr\rho^2=\fr{1}{2^N}(1+\sum_{k=1}^N ||\mathbf{s^{(k)}}||^2+\sum_{\{k_1,k_2\}}||\mathcal{T}^{\{k_1,k_2\}}||^2+\cdots+\sum_{\{k_1,k_2,\cdots,k_M\}}||\mathcal{T}^{\{k_1,k_2,\cdots,k_M\}}||^2$$
$$+\cdots+||\mathcal{T}^{(N)}||^2)=1 \eqno{(5)}$$

Any state $\rho=|\psi\ran\lan\psi|$ living in $d^2$ dimensional Hilbert space of operators acting on a $d$ dimensional Hilbert space of kets, can be expanded in the basis comprising $d^2-1$ generators of SU(d) and the identity operator. The set of coefficients in this expansion, namely $\{Tr(\rho\lambda_i)\}\;i=1,2,\cdots,d^2-1$ is a vector in $\mathbb{R}^{d^2-1}$ and is the Bloch vector of $\rho.$ The set of Bloch vectors and the set of density operators are in one to one correspondence with each other. The set of Bloch vectors for a given system forms a subspace of $\mathbb{R}^{d^2-1}$ denoted  $B(\mathbb{R}^{d^2-1}).$ The specification of this subspace for $d\geq 3$ is an open problem [25,26]. However, for pure states, following results are known [27]. $$\Arrowvert s \Arrowvert_2=\sqrt{\fr{d(d-1)}{2}}\eqno(6)$$ $$D_{r}(\mathbb{R}^{d^2-1})\subseteq B(\mathbb{R}^{d^2-1})\subseteq D_{R}(\mathbb{R}^{d^2-1})$$ where $D_r$ and $D_R$ are the balls of radii $r=\sqrt{\fr{d}{2(d-1)}}$ and $R=\sqrt{\fr{d(d-1)}{2}}$ respectively in $\mathbb{R}^{d^2-1}.$

We propose the following measure for a $N$-qubit pure state entanglement $$E_{\mathcal{T}}(|\psi\ran)=(||\mathcal{T}^{(N)}||-1) \eqno{(7)}$$

where $\mathcal{T}^{(N)}$ is given by Eq.(4) for ($M=N$) in Bloch representation of $\rho=|\psi\ran \lan\psi|$. The norm of the tensor $\mathcal{T}^{(N)}$ appearing in definition (7) is the Hilbert-Schmidt (Euclidean) norm $||\mathcal{T}^{(N)}||^2=(\mathcal{T}^{(N)},\mathcal{T}^{(N)})= \sum_{\alpha_{1}\alpha_{2}\cdots \alpha_{N}}t_{\alpha_{1}\alpha_{2}\cdots \alpha_{N}}^2$. Throughout this paper, by norm, we mean the Hilbert-Schmidt (Euclidean) norm. We comment on the normalization of $E_{\mathcal{T}}(|\psi\ran)$ below.\\

\bc

\textbf{III. $GHZ$ AND $W$ STATES}\\

\ec

Before proving various properties of $E_{\mathcal{T}}(|\psi\ran)$, we evaluate it for states in the $N$-qubit $GHZ$ or $W$ class. A general $N$-qubit $GHZ$ state is given by $$|\psi\ran=\sq{p}|000\cdots0\ran+\sq{1-p}|111\cdots1\ran; \;\;N\ge 2 \eqno{(8)}$$ 

A general element of $\mathcal{T}^{(N)}$ is given by $t_{i_1i_2\cdots i_N}=\lan\psi|\si_{i_1}\otimes\si_{i_2}\otimes\cdots\otimes\si_{i_N}|\psi\ran,\;i_k=1,2,3;\;k=1,2,\cdots ,N.$
The nonzero elements of $\mathcal{T}^{(N)}$ are $t_{11\cdots1}=2\sq{p(1-p)}$, $t_{33\cdots3}=p+(-1)^N(1-p)$.
Other nonzero elements of $\mathcal{T}^{(N)}$ are those with $2k\;\si_2$s and $(N-2k)\;\si_1$s, $k=0,1,\cdots,\lfloor \fr{N}{2} \rfloor$ where $\lfloor x \rfloor$ is the greatest integer less than or equal to $x$. (eg. for $N=3,\; t_{122}\; etc)$. These are equal to $(-1)^k \;2 \sq{p(1-p)}.$ This gives $$||\mathcal{T}^{(N)}||^2=4p(1-p)+(p+(-1)^N(1-p))^2+4p(1-p)\sum_{k=1}^{\lfloor \fr{N}{2} \rfloor}\binom{N}{2k}. \eqno{(9)}$$

Thus we get, for  $E_{\mathcal{T}}(|\psi\ran)$

$$E_{\mathcal{T}}(|\psi\ran)=||\mathcal{T}^{(N)}||-1 =\sq{4p(1-p)+(p+(-1)^N(1-p))^2+4p(1-p)\sum_{k=1}^{\lfloor \fr{N}{2} \rfloor}\binom{N}{2k}}-1 \eqno{(10)}$$

Eq.(8), with $N=2,$ represents a general two qubit entangled state in its Schmidt decomposition $$\arrowvert\psi\ran=\sq{p}\arrowvert00\ran+\sq{1-p}\arrowvert11\ran.$$ 
Thus Eq.(10) gives the entanglement in a two qubit pure state. Using Eq.(10) it is straightforward to see that $E_{\mathcal{T}}(|\psi\ran)$ for an arbitrary two qubit pure state is related to concurrence by
$$E_{\mathcal{T}}(|\psi\ran)=\sq{1+2C^2}-1,$$ where concurrence $C$ for such a state is $2\sq{p(1-p)}.$
   
Figure 1  plots $E_{\mathcal{T}}(\psi\ran)$ in Eq. (10) as a function of $p$ for $N=3$. For the $N$-qubit $GHZ$ (maximally entangled) state $p=1/2$, so that
$$R_N=E_{\mathcal{T}}(|GHZ\ran)=\sq{1+\fr{1}{4}(1+(-1)^N)^2+\sum_{k=1}^{\lfloor \fr{N}{2} \rfloor}\binom{N}{2k}}-1 \eqno{(11)}$$

\begin{figure}[!ht]
\begin{center}
\includegraphics[width=10cm,height=8cm]{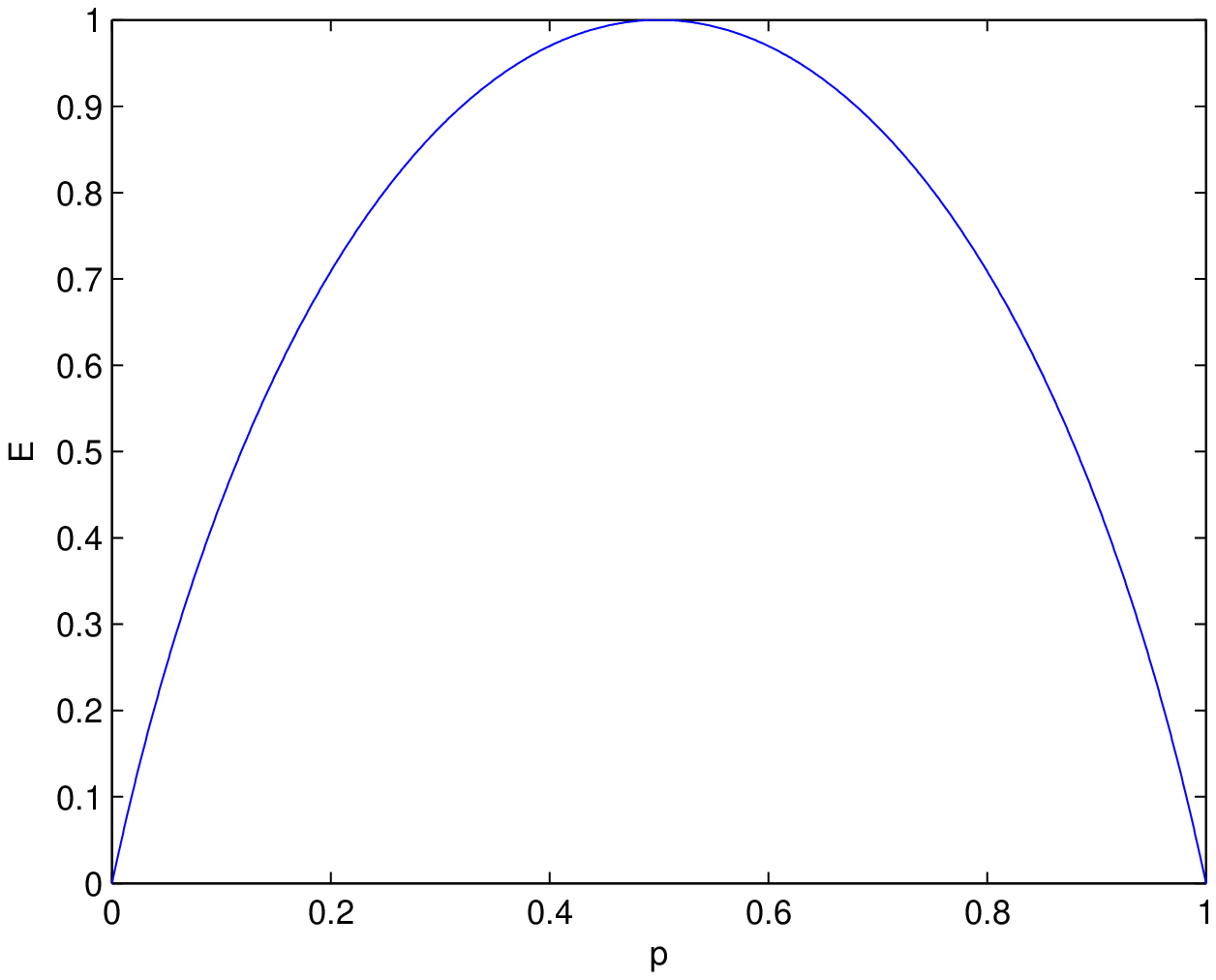}

FIG. 1. Variation of $E_{\mathcal{T}}(|GHZ\ran)$ for $N=3,$ express in units of $R_3,$ with parameter $p$.
\end{center}
\end{figure}

We see that, as a function of $N$, $E_{\mathcal{T}}(|GHZ\ran)$ increases as a polynomial of degree $\lfloor \fr{N}{2} \rfloor$. Figure 2 plots $E_{\mathcal{T}}(|GHZ\ran)$ as a function of $N$. $E_{\mathcal{T}}(|GHZ\ran)$ increases sharply with $N$ as expected. Note that $E_{\mathcal{T}}(|GHZ\ran)\ge 0$ for $GHZ$ class of states. Whenever appropriate, we normalize the entanglement of a $N$-qubit state $|\psi\ran$, $E_{\mathcal{T}}(|\psi\ran)$ by dividing by $R_N=E_{\mathcal{T}}(|GHZ\ran).$

\begin{figure}[!ht]
\begin{center}
\includegraphics[width=10cm,height=8cm]{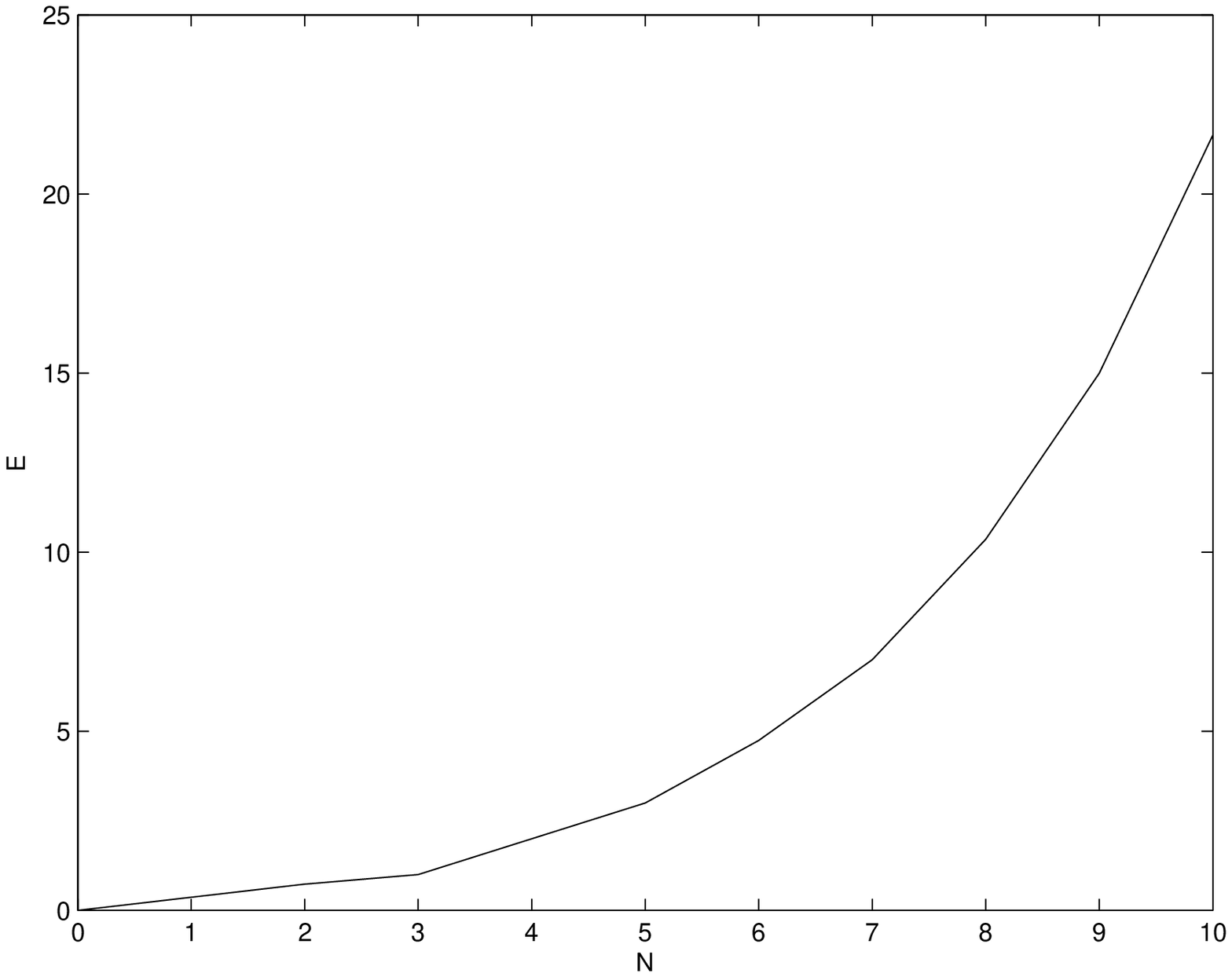}

FIG. 2. Variation of $E_{\mathcal{T}}(|GHZ\ran)$ with number of qubits $N$. 
\end{center}
\end{figure}

\vspace{.2in}

$N$-qubit W state is given by $$|W\ran=\fr{1}{\sq{N}}\sum_j|00\cdots 1_j 0 \cdots 00\ran;\; \;N \ge 3$$
where $j$th term has a single 1 at the $j$th bit. The state $|\widetilde{W}\ran=\otimes_{k=1}^N \si_1^{(k)} |W\ran$ is  given by $|\widetilde{W}\ran=\fr{1}{\sq{N}}\sum_j |11\cdots 0_j 1 \cdots 11\ran\; ;N\ge 3$ and has a single 0 at the $j$th bit. We note that $|\widetilde{W}\ran$ is locally unitarily connected to $|W\ran$ so that their entanglements must measure to the same value. The general element of $\mathcal{T}^{(N)}$ for the state $\rho=|W\ran \lan W|$ is 

$$t_{i_1i_2\cdots i_N}=\fr{1}{N}\sum_{j=1}^{N}\lan00\cdots 1_j\cdots00|\si_{i_1}\otimes\si_{i_2}\otimes\cdots\otimes\si_{i_N}|00\cdots 1_j\cdots00\ran$$

$$+\fr{1}{N}\sum_{j,l=1;j\ne l}^{N}\lan00\cdots1_j\cdots00|\si_{i_1}\otimes\si_{i_2}\otimes\cdots\otimes\si_{i_N}|00\cdots 1_l\cdots00\ran$$
Only the first term contributes to  $t_{33\cdots33}=-1$. Other nonzero elements have the form $t_{3\cdots31_j3\cdots 31_l3\cdots3}=\fr{2}{N}=t_{3\cdots32_j3\cdots 32_l3\cdots3}.$

There are $\binom{N}{2}$ elements of each of these two types, so that $$||\mathcal{T}^{(N)}||^2=1+2\Big(\fr{2}{N}\Big)^2 \binom{N}{2}=1+4\fr{N-1}{N}\eqno{(12)}$$

$$E_{\mathcal{T}}(|W\ran)=||\mathcal{T}^{(N)}||-1 =\sq{1+4\fr{N-1}{N}}-1 \eqno{(13)}$$

It is straightforward to check that $E_{\mathcal{T}}(|W\ran)=E_{\mathcal{T}}(|\widetilde{W}\ran)$ as expected. Note that $E_{\mathcal{T}}(|W\ran)\ge 0.$

Next we consider a superposition of $|W\ran$ and $|\widetilde{W}\ran$ states, 
$|\psi_{s,\phi}\ran=\sq{s}|W\ran+\sq{1-s}e^{i\phi}|\widetilde{W}\ran$.
It is clear that the entanglement of $|\psi_{s,\phi}\ran$ cannot depend on the relative phase $\phi$, as $|\psi_{s,\phi}\ran$ is invariant under the local unitary transformation $\{|0\ran,|1\ran\} \rightarrow \{|0\ran,e^{i\phi}|1\ran \}$ upto an overall phase factor. As we shall prove below, $E_{\mathcal{T}}$ is invariant under local unitary transformations. Figure 3 shows the entanglement of $|\psi_{s,\phi}\ran$ as a function of $s$, calculated using our measure.

\begin{figure}[!ht]
\begin{center}
\includegraphics[width=10cm,height=8cm]{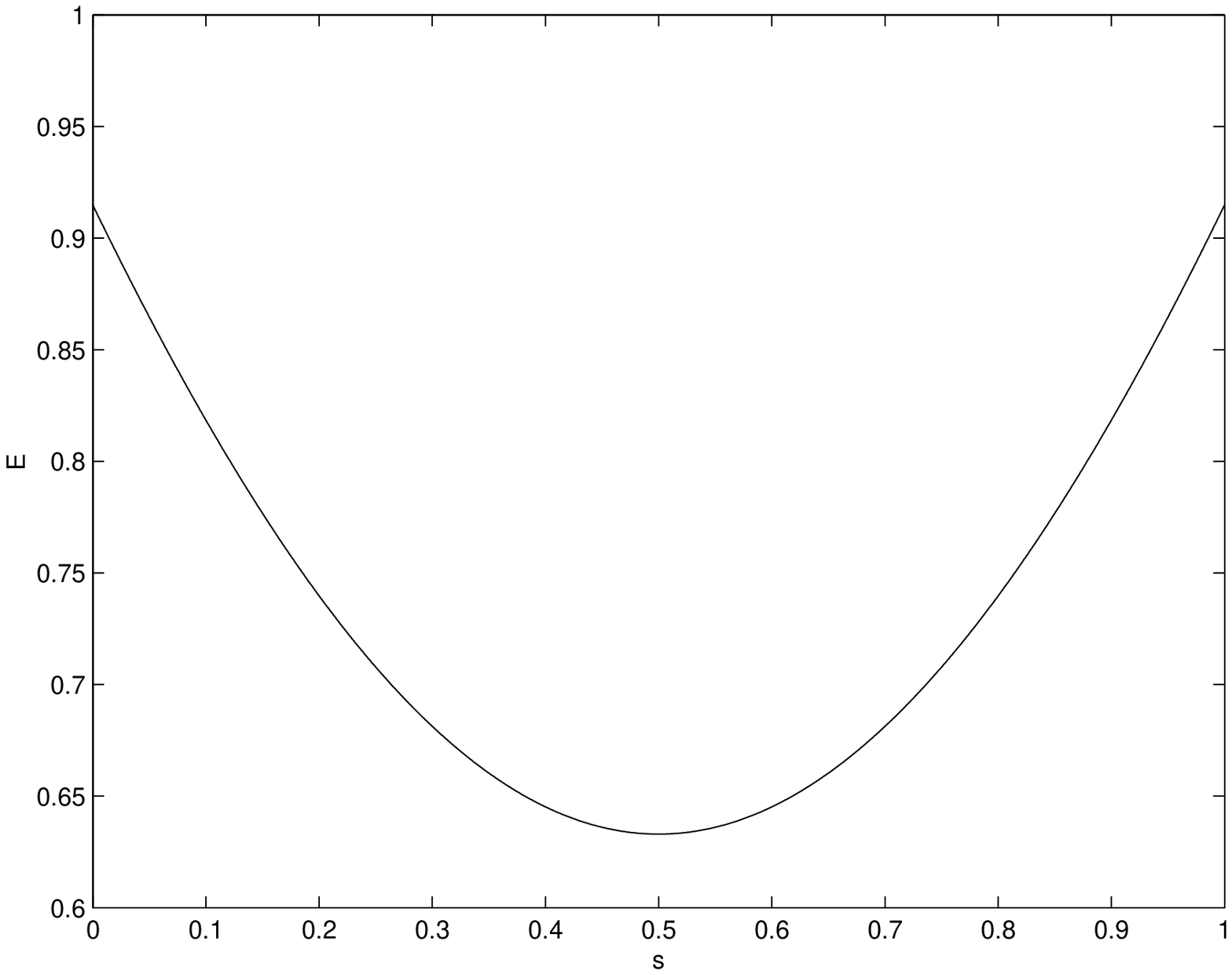}

FIG. 3. Variation of  $E_{\mathcal{T}}(|\psi_{s,\phi}\ran),$ expressed in units of $R_N,$ with the superposition parameter $s$.
\end{center}
\end{figure}

An important example of $W$ state and its generalizations is the one dimensional spin-$\fr{1}{2}$ Heisenberg antiferromagnet, on the lattice of size $N,$ with periodic boundary conditions, given by the Hamiltonian 
$$ H_{N}= \sum^{N}_{j=1}X_{j}X_{j+1}+ Y_{j}Y_{j+1}+Z_{j}Z_{j+1}\eqno{(14)}$$ where the subscripts are $mod \;N $ and $X,Y,Z$ denote Pauli operators $\si_x,\si_y,\si_z$ respectively. $H_N$ commutes with $S_z=\sum Z_j$, so the eigenstate of $H_N$ is a superposition of basis vectors $|b_1\cdots b_n\ran$ with $s$ of $b_1\cdots b_N$ are ones and $N-s$ are zeros for some fixed $0 \le s \le N.$ When $s=1$, the translational invariance of $H_N$ implies that the eigenstates are $$|\psi^{(k)}_N\ran=\fr{1}{\sq{N}}\sum_{j=0}^{N-1} e^{ikj}|00\cdots 1_j0\cdots 0\ran \eqno{(15)}$$

where the $j$th summand has a single 1 at $j$th bit just like $W$ state and the wave number $k=\fr{2 \pi m}{N}$ for some integer $0\le m \le N-1$. The state $|\psi^{(k)}_N\ran$ is locally unitarily transformed to the $W$ state so that it has the same value of $E_{\mathcal{T}}(|W\ran)$ or $E_{\mathcal{T}}(\widetilde{W}\ran).$

For $ s \ge 2$ the eigenstates of $H_N$ have the form $$|\psi_N(s)\ran=\fr{1}{\sq{\binom{N}{s}}}\sum_{\{j_1 \cdots j_s\}} |00\cdots 1_j0 \cdots 1_{j_s} 0\ran, \eqno{(16)}$$

where 1 occurs at $j_1\cdots j_s$, $\{j_1\cdots j_s\}\subseteq \mathcal{N}=\{1,2,\cdots,N\}$ and can be chosen in $\binom{N}{s}$ ways. We see that for $|\psi_N(s)\ran$, $t_{33\cdots3}=1.$ For even $N$, $t_{1\cdots12\cdots23\cdots3}$ with $x$ $1$ s and $y$ $2$ s, corresponding to the average of $x$ $\si_x$ s, $y$ $\si_y$ s and $N-x-y$ $\si_z$ s, we get, for even $x$ and even $y$,
$$t_{1\cdots12\cdots23\cdots3}=\Bigg[2\binom{x}{\fr{x}{2}}\binom{y}{\fr{y}{2}}-\binom{x+y}{\fr{(x+y)}{2}}\Bigg]\binom{N-x-y}{s-\fr{(x+y)}{2}}.$$  
Since $|\psi_N(s)\ran$ is a symmetric state, any permutation of its indices does not change the value of an element of $\mathcal{T}^{(N)}$ [24], so that,
$${\Arrowvert\mathcal{T}^{(N)}_{|\psi_{N}(s)\ran}\Arrowvert}^2=1+\fr{1}{{\binom{N}{s}}^2}\Bigg[
\sum_{\begin{subarray}{I} 
  \hskip  .1cm  {x+y=2}\\
 \hskip  .1cm   {x,y\; even}
\end{subarray}} ^{2s} {\Bigg[2\binom{x}{\fr{x}{2}}\binom{y}{\fr{y}{2}}-\binom{x+y}{\fr{(x+y)}{2}}\Bigg]}^2 {\binom{N-x-y}{s-\fr{(x+y)}{2}}}^2 \binom{N}{x} 
\binom{N-x}{y}\Bigg]$$
Figure 4 shows the variation of $E_{\mathcal{T}}(|\psi_{N}(s)\ran)=\Arrowvert{\mathcal{T}}_{|\psi_{N}(s)\ran}^{(N)}\Arrowvert
-1$ with $s$. We see that it is maximum at $s=\fr{N}{2}$, which is the ground state of $H_N$, as expected. Note that 
$E_{\mathcal{T}}(|\psi_{N}(\fr{N}{2})\ran)$ for the ground state $(s=\fr{N}{2})$ rises far more rapidly than the entanglement of the $N$-qubit $GHZ$ state $R_{N}=E_{\mathcal{T}}(|GHZ\ran),$ (Eq.(11)), with the number of spins (qubits) $N.$ This can be understood by noting that $|\psi_{N}(\fr{N}{2})\ran$ for $s=\fr{N}{2}$ can be written as a superposition of $\fr{1}{2}\binom{N}{\fr{N}{2}}$ $N$-qubit $GHZ$ states. For example, $|\psi_{4}(2)\ran$ can be written as the superposition of three 4-qubit $GHZ$ states,
$$|\psi_{4}(2)\ran=\fr{1}{\sq{3}}\Big[\fr{1}{\sq{2}}(\arrowvert 0011\ran+\arrowvert1100\ran)+ \fr{1}{\sq{2}}
(\arrowvert 0101\ran+\arrowvert1010\ran)+\fr{1}{\sq{2}}(\arrowvert 1001\ran+\arrowvert0110\ran)\Big].$$
As $N$ increases, initially $E_{\mathcal{T}}(|\psi_{N}(\fr{N}{2})\ran)$ is comparable to $R_{N},$ but after $N=16$ the ratio 
 $\fr{E_{\mathcal{T}}(|\psi_{N}(\fr{N}{2})\ran)}{R_{N}}$ increases very rapidly, reaching $10^7$ for $100$ qubits. Also, as $N$ increases, $E_{\mathcal{T}}(|\psi_{N}(s)\ran)$ falls off more rapidly as $s$ deviates from $\fr{N}{2}$. We are presently trying to understand this behavior.

\begin{figure}[!ht]
\begin{center}
\includegraphics[width=10cm,height=8cm]{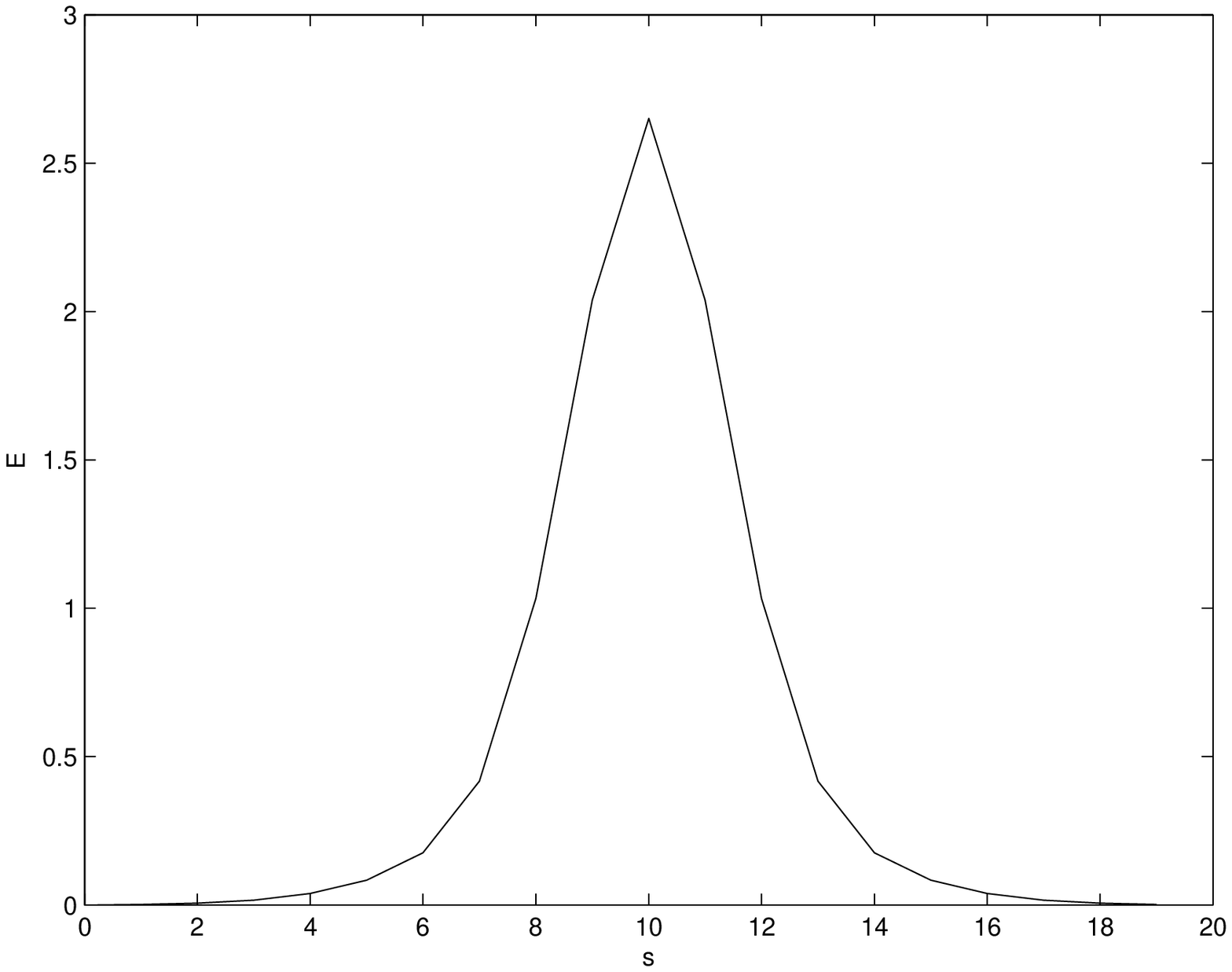}

FIG. 4a 
\end{center}
\end{figure}

\begin{figure}[!ht]
\begin{center}
\includegraphics[width=10cm,height=8cm]{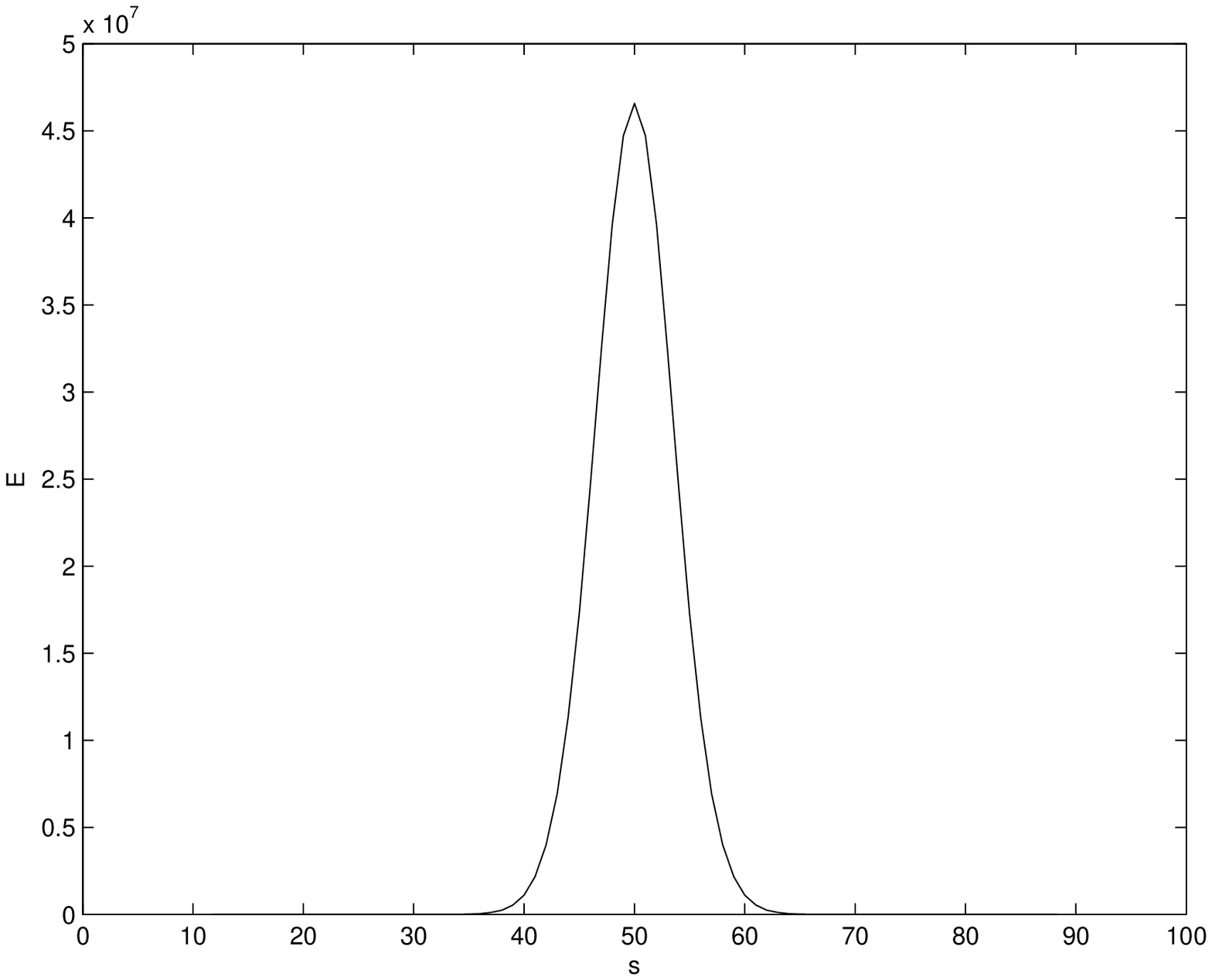}

FIG. 4b\\

FIG. 4. Variation of $E_{\mathcal{T}}(|\psi_{N}(s)\ran),$ in units of $R_N$ with $s$, for (a) $N=20$ and (b) $N=100$ (see text).
 
\end{center}
\end{figure}

Finally, in this section, we consider the superpositions of $W$ and $GHZ$ states,
$$|\psi_{W+GHZ}(s,\phi)\ran=\sq{s}|GHZ\ran+\sq{1-s}\; e^{i\phi}|W\ran, \eqno{(17)}$$ also considered in [28]. For three qubits, $N=3$, a direct calculation gives, for this state, $${\Arrowvert\mathcal{T}^{(N)}\Arrowvert}^2\;=\;4s^2+6s(1-s)+\fr{11}{3}(s-1)^2; \;\;(0\leq s\leq 1) \eqno{(18)}$$ 
$$E_{\mathcal{T}}(|\psi_{W+GHZ}(s,\phi)\ran)=\Arrowvert{\mathcal{T}}_{|\psi_{N}(s)\ran}^{N}\Arrowvert
-1$$
which coincides with the corresponding values of $W\;(s=0)$ and $GHZ\;(s=1)$ states. Note that $E_{\mathcal{T}}(|\psi_{W+GHZ}(s,\phi)\ran)$
is independent of the phase $\phi,$ in contrast to the entanglement measure used in [28]. Figure 5 shows the dependence of 
$E_{\mathcal{T}}(|\psi_{W+GHZ}(s,\phi)\ran)$ on $s$.\\

\begin{figure}[!ht]
\begin{center}
\includegraphics[width=10cm,height=8cm]{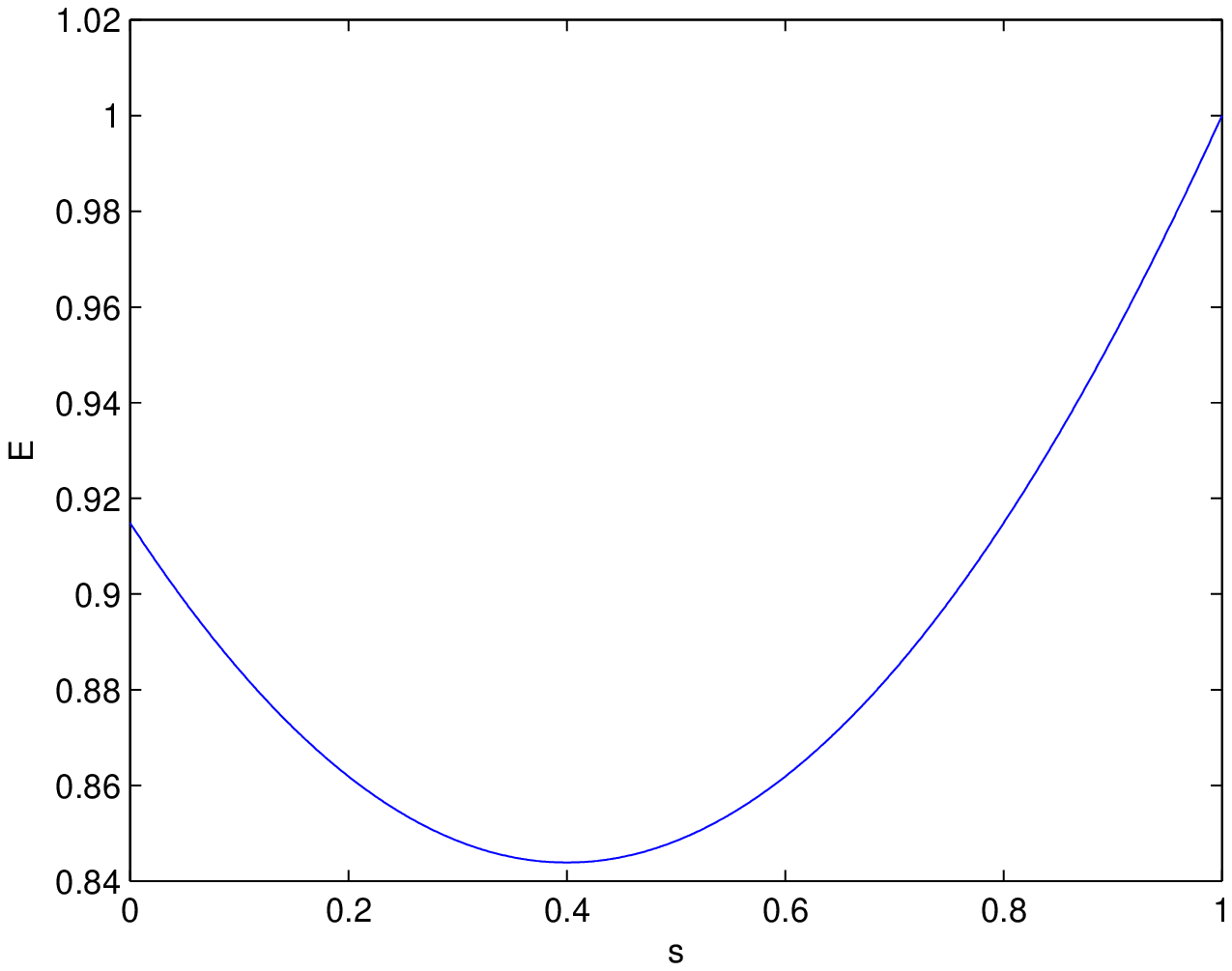}

FIG. 5. Variation of $E_{\mathcal{T}}(|\psi_{W+GHZ}(s,\phi)\ran)$, expressed in units of $R_N$, with the superposition parameter $s$, for $N=3$.
\end{center}
\end{figure}

\bc

\textbf{ IV. PROPERTIES OF $E_{\mathcal{T}}(|\psi\ran)$}\\

\ec
To be a valid entanglement measure, $E_{\mathcal{T}}(|\psi\ran)$ must have the following properties [29,30].

(a) (i) {\it Positivity} : $E_{\mathcal{T}}(|\psi\ran)\ge 0$ for all $N$-qubit pure state $|\psi\ran$.
      (ii) {\it Discriminance}: $E_{\mathcal{T}}(|\psi\ran)=0$ if and only if $|\psi\ran$ is separable (product) state. 
  
(b) {\it $LU$ invariance} : $E_{\mathcal{T}}(|\psi\ran)$ is  invariant under local unitary operations.

(c) {\it Monotonicity} : local operators and classical communication ($LOCC$) do not increase the expectation value of $E_{\mathcal{T}}(|\psi\ran)$.

 We prove the above properties  for $E_{\mathcal{T}}(|\psi\ran)$. We also prove the following additional properties for $E_{\mathcal{T}}(|\psi\ran)$.
 
(d) {\it continuity} $||(|\psi\ran\lan\psi|-|\phi\ran\lan\phi|)|| \rightarrow 0 \Rightarrow \Big{|}E(|\psi\ran)-E(|\phi\ran)\Big{|}\rightarrow 0 $.
  
(e) {\it superadditivity} $E_{\mathcal{T}}(|\psi\ran \otimes |\phi\ran )\ge E_{\mathcal{T}}(|\psi\ran)+ E_{\mathcal{T}}(|\phi\ran).$\\

We need the following result which we have proved in [24].\\

\textit{ Proposition 0 }: A  pure $N$-partite quantum state is fully separable (product state) if and only if   $$\mathcal{T}^{(N)}=\mathbf{s}^{(1)}\circ \mathbf{s}^{(2)} \circ \dots \circ \mathbf{s}^{(N)},\eqno{(19)}$$ 
     where $ \mathbf{s}^{(k)}$ is the Bloch vector of $k$th subsystem reduced density matrix.

The symbol $\circ$ stands for  the outer product of vectors defined as follows.
  
Let $\mathbf{u}^{(1)},\mathbf{u}^{(2)},\dots,\mathbf{u}^{(M)}$ be vectors in $\mathbb{R}^{d_1^2-1},\mathbb{R}^{d_2^2-1},\cdots,\mathbb{R}^{d_M^2-1}.$ 
The outer product $\mathbf{u}^{(1)}\circ \mathbf{u}^{(2)} \circ \dots \circ \mathbf{u}^{(M)}$ is a tensor of order $M$, (M-way array), defined by

$ t_{i_1 i_2\cdots i_M}=\mathbf{u}^{(1)}_{i_1} \mathbf{u}^{(2)}_{i_2} \dots  \mathbf{u}^{(M)}_{i_M};\; 1\le i_k \le d_k^2-1,\; k=1,2,\cdots,M.$

\textit{ Proposition 1 }: Let $|\psi\ran$ be a $N$-qubit pure state. Then, $||\mathcal{T}^{(N)}_{\psi}||=1$ if and only if $|\psi\ran$ is a separable (product) state.

\textit{Proof.} By proposition 0, $|\psi\ran$ is separable (product) if and only if $$\mathcal{T}^{(N)}=\mathbf{s}^{(1)}\circ \mathbf{s}^{(2)} \circ \dots \circ \mathbf{s}^{(N)},$$ 

As shown in [31,32 ], $$( \bigcirc_{k=1}^N s^{(k)},\bigcirc_{k=1}^N s^{(k)})=\Pi_{k=1}^N ( s^{(k)},s^{(k)}),\eqno{(20)}$$

where $(,)$ denotes the scaler product. This immediately gives, for qubits, $$||\mathcal{T}^{(N)}||^2=( \mathcal{T}^{(N)},\mathcal{T}^{(N)}) =\Pi_{k=1}^N ( s^{(k)},s^{(k)})= \Pi_k ||s^{(k)}||^2=1$$ 

Proposition 1 immediately gives

\textit{ Proposition 2 } Let $|\psi\ran$ a $N$-qubit pure state. Then $E_{\mathcal{T}}(|\psi\ran)= 0$ if and only if $|\psi\ran$ is a product state.

\textit{Proposition 3 }: Let $|\psi\ran$ be a $N$-qubit pure state. Then $ ||\mathcal{T}^{(N)}||\ge 1.$

It is instructive to show this result by direct computation of $\mathcal{T}^{(N)}$ for the case of two and three qubit states. First, consider a general two qubit state $$|\psi\ran=a_1|00\ran+a_2|01\ran+a_3|10\ran+a_4|11\ran,\; \sum_k|a_k|^2=1$$

by direct computation we get $$||\mathcal{T}^{(2)}||^2=1+8(|a_2a_3|-|a_1a_4|)^2 \ge 1 \eqno{(21)}$$

This means, via proposition 2 that $|\psi\ran$ is a product state if $|a_2a_3|=|a_1a_4|.$
Next, consider a three qubit state in the general Schmidt form [33] $$|\psi\ran=\la_0|000\ran+\la_1e^{i\phi}|100\ran+\la_2|101\ran+\la_3|110\ran+\la_4|111\ran \eqno{(22)}$$

where $\la_i \ge 0, \; i=0,1,2,3,4$ and $\sum_i \la_i^2=1$

By direct calculation of $||\mathcal{T}^{(3)}||$ we get
$$||\mathcal{T}^{(3)}||^2 \ge1+12\la_0^2\la_4^2+8\la_0^2\la_2^2+8\la_0^2\la_3^2+8(\la_1\la_4-\la_2\la_3)^2 \ge 1 \eqno{(23)}$$

Here the conditions for product state become $(\la_1\la_4=\la_2\la_3 $ and $ \la_0=0$).
We now prove proposition 4 for a general $N$-qubit state $|\psi\ran.$

If $|\psi\ran$ is not a product of $N$ single qubit states (i.e. $|\psi\ran$ is not $N$-separable) then it is $N-k$ separable $k=2,3,\cdots,N-1$. Viewing the $N$-qubit system as a system comprising $N-k$ qubits, each with Hilbert space of of dimension 2 and $k$ entangled qubits with the Hilbert space of dimension $2^k$, we can apply proposition 0 to this separable system of $N-k+1$ parts in the state $|\psi\ran.$ We get $ \mathcal{T}^{(N)}_{|\psi\ran}= \mathbf(s^{(1)})\circ\mathbf(s^{(2)})\circ \cdots\mathbf(s^{(N-k)})\circ\mathbf(s^{(N-k+1)})$

This implies, as in proposition 1, via Eq. (20) and Eq. (6) that $$||\mathcal{T}^{(N)}_{|\psi\ran}||=\Pi_{i-1}^{N-k+1} ||\mathbf(s^{(i)})||^2=\fr{d_k(d_k-1)}{2} > 1 \; (d_k=2^k).\eqno{(24)}$$
If $k=N$ we attach an ancilla qubit in an arbitrary state $|\phi\ran$ and apply proposition 0 to $N+1$ qubit system in the state $|\psi\ran\otimes|\phi\ran$ where $|\psi\ran$ is the $N$-qubit entangled state. This result, combined with proposition 1,  completes the proof.

Proposition 4 immediately gives\\

\textit{Proposition 5} : $E_{\mathcal{T}}(|\psi\ran)\ge 0.$\\

We now prove that $E_{\mathcal{T}}(|\psi\ran)$ is nonincreasing under local operations and classical communication. Any such local action can be decomposed into four basic kinds of operations [34] (i) appending an ancillary system not entangled with the state of original system, (ii) performing a unitary transformation, (iii) performing measurements, and (iv) throwing away, i.e. tracing out, part of the system. It is clear that appending ancilla cannot change  
$\Arrowvert\mathcal{T}^{(N)}\Arrowvert.$ We prove that  $E_{\mathcal{T}}(|\psi\ran)$ does not increase under the remaining three local operations.

\textit{Proposition 6} : Let $U_i \; (i=1,2,\cdots,N)$ be a local unitary operator acting on the Hilbert space of $i$th subsystem  $\mathcal{H}^{(i)}$.

Let $$\rho=(\otimes_{i=1}^N U_i)\rho'(\otimes_{i=1}^N U_i^{\dag})  \eqno{(25)}$$
for  density operators $\rho$ and $\rho'$ acting on $\mathcal{H}=\otimes_{i=1}^{N}\mathcal{H}^{(i)}$ and let $\mathcal{T}^{(N)}$ and $\mathcal{T'}^{(N)}$ denote the $N$ partite correlation tensors for $\rho$ and $\rho'$  respectively. Then, 

 $||\mathcal{T'}^{(N)}||=||\mathcal{T}^{(N)}||,$ so that $E_{\mathcal{T}}(\rho)=E_{\mathcal{T}}(\rho')$\\

 \textit{Proof.} Let $U$ denote a one qubit unitary operator, then it is straightforward to show that $U \si_{\alpha}U^{\dag}=\sum_{\beta} O_{\alpha \beta}\si_{\beta}$
 
 where $[O_{\alpha \beta}]$ is a real matrix satisfying $O O^T =I= O^T O$. 
 It is an element of the rotation group $O(3)$. Now consider $$t'_{i_1i_2\cdots i_N}=Tr(\rho'\si_{i_1}\otimes\si_{i_2}\otimes\cdots \otimes\si_{i_N})$$
 $$=Tr \big{(}\rho(\otimes_{i=1}^N U_i) \si_{i_1}\otimes\si_{i_2}\otimes\cdots \otimes\si_{i_N}(\otimes_{i=1}^N U_i^{\dag})\big{)}$$
 $$=Tr(\rho U_1\si_{i_1}U_1^{\dag}\otimes U_2\si_{i_2}U_2^{\dag}\otimes\cdots \otimes U_N\si_{i_N}U_N^{\dag})$$
 
 $$=\sum_{\alpha_1\cdots\alpha_N}Tr(\rho\si_{\alpha_1}\otimes\si_{\alpha_2}\otimes\cdots \otimes\si_{\alpha_N})O^{(1)}_{i_1\alpha_1}O^{(2)}_{i_2\alpha_2}\cdots O^{(N)}_{i_N\alpha_N}$$
 $$=\sum_{\alpha_1\cdots\alpha_N}t_{\alpha_1\cdots\alpha_N}O^{(1)}_{i_1\alpha_1}O^{(2)}_{i_2\alpha_2}\cdots O^{(N)}_{i_N\alpha_N}$$
 $$=(\mathcal{T}^{(N)}\times_1 O^{(1)}\times_2 O^{(2)}\cdots \times_N O^{(N)})_{i_1i_2\cdots i_N}$$

where $\times_k $ is the $k$-mode product of a tensor $\mathcal{T}^{(N)}\in \mathbb{R}^{3\times 3 \times \cdots 3}$ by the orthogonal matrix $O^{(k)}\in \mathbb{R}^{3\times 3}$ [31,32,35]. Therefore, 

$$\mathcal{T'}^{(N)}=\mathcal{T}^{(N)}\times_1 O^{(1)}\times_2 O^{(2)}\cdots \times_N O^{(N)}$$

By proposition 3.12 in [31] we get $$||\mathcal{T'}^{(N)}||= ||\mathcal{T}^{(N)}\times_1 O^{(1)}\times_2 O^{(2)}\cdots \times_N O^{(N)}||=||\mathcal{T}^{(N)} ||$$ \hfill $\blacksquare$\\

\textit{Proposition 7} : If a multipartite pure state $|\psi\ran$ is subjected to a local measurement on the $k$th qubit giving outcomes $i_k$ with probabilities $p_{i_k}$ and leaving residual $N$-qubit pure state $|\phi_{i_k}\ran$ then the expected entanglement $\sum_{i_k} p_{i_k} E_{\mathcal{T}}(|\phi_{i_k}\ran)$ of residual state is not greater then $ E_{\mathcal{T}}(|\psi\ran)$.

$$\sum_{i_k} p_{i_k} E_{\mathcal{T}}(|\phi_{i_k}\ran) \le E_{\mathcal{T}}(|\psi\ran). \eqno{(26)}$$

\textit{Proof.} Local measurements can be expressed as the tensor product matrix $\bar{D}=\bar{D}^{(1)}\otimes\bar{D}^{(2)}\otimes\cdots \otimes \bar{D}^{(N)}$  on the expanded coherence vector $\mathcal{T}$ [36]. The expanded coherence vector $\mathcal{T}$ is the extended correlation tensor $\mathcal{T}$ (defined below) viewed as a vector in the real space of appropriate dimension. The extended correlation tensor $\mathcal{T}$ is defined by the equation
$$\rho= \fr{1}{2^N}\sum_{i_1i_2\cdots i_N=0}^3 \mathcal{T}_{i_1i_2\cdots i_N} \si_{i_1}\otimes \si_{i_2}\otimes\cdots\otimes \si_{i_N}  \eqno{(27)}$$
where $\si_{i_k} \in \{I,\si_x,\si_y,\si_z\}$ is the $i_k$th local Pauli operator on the $k$th qubit $(\si_0 =I)$ and the real coefficients $\mathcal{T}_{i_1i_2\cdots i_N} $ are the components of the extended correlation tensor $\mathcal{T}.$  Eq. (2) and Eq.(27) are equivalent with $\mathcal{T}_{000\cdots 0}=1 $ , $\mathcal{T}_{i_100\cdots 0}= s^{(1)}_{i_1},\; \cdots $, $\mathcal{T}_{i_1i_2\cdots i_M 00\cdots 0} = \mathcal{T}^{\{1,2,\cdots M\}}_{i_1i_2\cdots i_M}, \cdots $
and  $\mathcal{T}_{i_1i_2\cdots i_N} =\mathcal{T}^{(N)}_{i_1i_2\cdots i_N} ,\; i_1, i_2,\cdots ,i_N \ne 0 .$ $\bar{D}^{(k)};\; k=1,2,\cdots n$ are $4\times 4$ matrices. Without losing generality, we can assume the local measurements to be POVMs, in which case $\bar{D}^{(k)}= diag(1,D^{(k)})$ and the $3  \times 3$ matrix $D^{(k)}$ is contractive $D^{(k)T} D^{(k)} \le I$ [36]. The local POVMs acting on a $N$-qubit state $\rho$ corresponds to the map 
$\rho\longmapsto\mathcal{M}(\rho)$ given by
$$\mathcal{M}(\rho)=\sum_{i_1i_2\cdots i_N}L^{(1)}_{i_1}\otimes L^{(2)}_{i_2}\otimes \cdots \otimes L^{(N)}_{i_N}\rho L^{(1)\dag}_{i_1}\otimes L^{(2)\dag}_{i_2}\otimes \cdots \otimes L^{(N)\dag}_{i_N}$$
where $L^{(k)}_{i_k}$ are the linear, positive, trace preserving operators satisfying $\sum_{i_k}L^{(k)\dag}_{i_k}L^{(k)}_{i_k}=I$ and $[L^{(k)}_{i_k},L^{(k)\dag}_{i_k}]=0.$ The resulting correlation tensor of $\mathcal{M}(\rho)$ can be written as 

$$\mathcal{T'}^{(N)}=\mathcal{T}^{(N)}\times_1 D^{(1)}\times_2 D^{(2)}\cdots \times_N D^{(N)}$$
where $D^{(k)}$ is $3\times3$ matrix and $D^{(k)T} D^{(k)} \le I.$

Action of POVM on $k$th qubit corresponds to the map $\mathcal{M}_{k}(|\psi\ran\lan\psi|)=\sum_{i_k} M_{i_k}\rho M_{i_k}^{\dag}$ where $M_{i_k}=I \otimes \cdots L^{(k)}_{i_k}\otimes \cdots I$ , $\sum_{i_k}L^{(k)\dag}_{i_k}L^{(k)}_{i_k}=I$ and $[L^{(k)}_{i_k},L^{(k)\dag}_{i_k}]=0$
with the resulting mixed state $\sum_{i_k}p_{i_k}|\phi_{i_k}\ran\lan\phi_{i_k}|,$ where $|\phi_{i_k}\ran$ is the $N$-qubit pure state which results after the the outcome $i_k$ with probability $p_{i_k}.$

The average entanglement of this state is $$\sum_{i_k} p_{i_k} E_{\mathcal{T}}(|\phi_{i_k}\ran\lan\phi_{i_k}|)=\sum_{i_k}p_{i_k} ||\mathcal{T}^{(N)}_{|\phi_{i_k}\ran}||-1$$
$$=\sum_{i_k} p_{i_k}||\mathcal{T}^{(N)}_{|\psi\ran}\times_k D^{(k)}||-1$$
$$\sum_{i_k}p_{i_k}||D^{(k)}T_{(k)}(|\psi\ran)||-1$$
where, by proposition 3.7 in [31], $ D^{(k)}T_{(k)}(|\psi\ran)$ is the $k$th matrix unfolding [24] of $\mathcal{T}^{(N)}_{|\psi\ran}\times_k D^{(k)}.$
Therefore, from the definition of the Euclidean norm of a matrix, $||A||=\sq{Tr(A A^{\dag})}$  [37] we get
$$\sum_{i_k} p_{i_k}  E_{\mathcal{T}}(|\phi_{i_k}\ran\lan\phi_{i_k}|)=\sum_{i_k} p_{i_k}\big{[}Tr\big{(}D^{(k)}T_{(k)}(|\psi\ran)T_{(k)}^{\dag}(|\psi\ran)D^{(k)T}\big{)}\big{]}^{\fr{1}{2}}-1$$

$$=\sum_{i_k} p_{i_k} \big{[}Tr \big{(} D^{(k)T}D^{(k)}T_{(k)}(|\psi\ran)T_{(k)}^{\dag}(|\psi\ran)\big{)}\big{]}^{\fr{1}{2}}-1$$

 $$\le \sum_{i_k}p_{i_k}\sq{Tr\big{(}T_{(k)}(|\psi\ran)T_{(k)}^{\dag}(|\psi\ran)\big{)}}-1$$
 $$=||\mathcal{T}^{(N)}_{|\psi\ran}||-1 = E_{\mathcal{T}}(|\psi\ran)$$ 
 because $D^{(k)T} D^{(k)} \le I$, and $\sum_{i_k}p_{i_k} = 1.$ We have also used the fact that Euclidean norm of a tensor equals that of any of its matrix unfoldings. \hfill $\blacksquare$\\
 
 As an example, we consider the $4-$qubit state [38]
 $$|\psi\ran_{ABCD}=\fr{1}{\sq{6}}(|0000\ran +|0011\ran +|0101\ran +|0110\ran +|1010\ran +|1111\ran). \eqno{(28)}$$
 A POVM $\{A_1,A_2\}$ is performed on the subsystem $A$, which has the form $A_1=U_1 diag\{\alpha, \beta \}V$
  and $A_2= U_2 diag\{\sq{1-\alpha^2}, \sq{1-\beta^2} \}V.$  Due to LU invariance of    $\Arrowvert\mathcal{T}^{(N)}\Arrowvert$ we need only consider the diagonal matrices in which the parameters are chosen to be $\alpha=0.9$ and $\beta=0.2.$ After the POVM, two outcomes $|\phi_1\ran=A_1|\psi\ran/\sq{p_1}$ and $|\phi_2\ran=A_2|\psi\ran/\sq{p_2}$ are obtained, with the probabilities as $p_1=0.5533$ and $p_2=0.4467.$ We find 
 $$E_{\mathcal{T}}(|\psi\ran)=0.7802,\;E_{\mathcal{T}}(|\phi_1\ran)=0.0725/p_1,\;E_{\mathcal{T}}(|\phi_2\ran)=0.0436/p_2.$$ This gives,$$E_{\mathcal{T}}(|\psi\ran)-[p_1 E_{\mathcal{T}}(|\phi_1\ran)+p_2 E_{\mathcal{T}}(|\phi_2\ran)]=0.6641\;>\;0.$$ This is to be contrasted with the similar calculation in [38], with the same state $|\psi\ran$ in Eq.(28) and the same POVM given above. \\ 
 
\textit{Proposition 8 :} Let $|\psi\ran$ be an $N$-qubit pure state. Let $\rho$ denote the reduced density matrix after tracing out one qubit from the state $|\psi\ran$. Then $$|| \mathcal{T}^{(N-1)}_{\rho}|| \le ||\mathcal{T}^{(N)}_{|\psi\ran}||$$
with equality only when $|\psi\ran=|\phi\ran\otimes|\chi\ran$ where $|\chi\ran$ is the state of the qubit which is traced out.\\

\textit{Proof.} we prove this for a special case whose generalization is straightforward. Let $|\psi\ran=a|b_1\cdots b_N\ran+b|b'_1\cdots b'_N\ran;\; |a|^2+|b|^2=1.$

Here $|b_i\ran$ and $|b'_i\ran$ are the eigenstates of $\si_z^{(i)}$ operating on the $i$th qubit. Now consider set of  $N$-fold tensor products of qubit operators $\{\si_{\alpha}\},\; \alpha=1,2,3$, namely  $S=\{\si_{\alpha_1}\otimes\si_{\alpha_2}\otimes \cdots \otimes \si_{\alpha_N}\},\; \alpha_1\cdots \alpha_N =1,2,3.$
Choosing $\alpha_1,\cdots, \alpha_N =3$ we get $\si_{3}\otimes\si_{3}\otimes \cdots\otimes \si_{3}|b_1\cdots b_N\ran=\pm |b_1\cdots b_N\ran.$
We can choose an operator from $S$, denoted $B$, such that $B|b_1\cdots b_N\ran = \pm |b'_1\cdots b'_N\ran.$
If $B$ contains $q \le N \; \si_x$ operators we can replace $k \le q$ of them by $\si_y$ operators. We denote the resulting tensor product operator by $B_k \; (B_0=B)$. We have, $B_k|b_1\cdots b_N\ran = \pm (i)^k |b'_1\cdots b'_N\ran.$
Then, $$\lan b_1\cdots b_N\arrowvert\si_3\otimes\cdots\otimes\si_3\arrowvert b_1\cdots b_N\ran = \pm 1 = \lan b^{\prime}_1\cdots b^{\prime}_N\arrowvert\si_3\otimes\cdots\otimes\si_3\arrowvert b^{\prime}_1\cdots b^{\prime}_N\ran$$
$$\lan b_1\cdots b_N\arrowvert B \arrowvert b^{\prime}_1\cdots b^{\prime}_N\ran = \pm 1 = \lan b^{\prime}_1\cdots b^{\prime}_N\arrowvert B \arrowvert b_1\cdots b_N\ran$$
$$\lan b^{\prime}_1\cdots b^{\prime}_N\arrowvert B_k \arrowvert b_1\cdots b_N\ran = \pm (i)^{k}$$
$$\lan b_1\cdots b_N\arrowvert B_k \arrowvert b^{\prime}_1\cdots b^{\prime}_N\ran = \pm (-i)^{k}$$
Now,$$ t_{\alpha_1\cdots\alpha_N} = \lan \psi\arrowvert\si_{\alpha_1}\otimes\cdots\otimes\si_{\alpha_N}\arrowvert\psi\ran$$
$$ = \arrowvert a\arrowvert^2 \lan b_1\cdots b_N\arrowvert\si_{\alpha_1}\otimes\cdots\otimes\si_{\alpha_N}\arrowvert b_1\cdots b_N\ran+|b|^2\lan b^{\prime}_1 \cdots b^{\prime}_N|\si_{\alpha_1}\otimes\cdots\otimes\si_{\alpha_N}|b^{\prime}_1 \cdots b^{\prime}_N\ran$$
$$ +a^* b \lan b_1\cdots b_N\arrowvert\si_{\alpha_1}\otimes\cdots\otimes\si_{\alpha_N}|b^{\prime}_1 \cdots b^{\prime}_N\ran+a b^* \lan b^{\prime}_1 \cdots b^{\prime}_N|\si_{\alpha_1}\otimes\cdots\otimes\si_{\alpha_N}\arrowvert b_1\cdots b_N\ran$$
The nonzero elements of $t_{\alpha_1 \cdots \alpha_N} $ are $t_{33\cdots 3}=\pm |a|^2 \pm |b|^2$,
$ t_B=\pm a b^* \pm a^* b= \pm 2 |a| |b| cos (\phi_a-\phi_b)$, 

 \begin{displaymath}
t_{B_k}=\pm (i)^k a b^* \pm (-i)^k a^* b 
 =\left\{ \begin{array}{ll}
 \pm 2 |a|\; |b| cos (\phi_a-\phi_b) & \textrm{ if $k$ is even} \\

\pm 2 |a|\; |b| sin (\phi_a-\phi_b) & \textrm{ if $k$ is odd} \\
\end{array} \right.
\end{displaymath}
We get $\sum _{k=0}^q \binom{q}{2k}$ elements with $cos (\phi_a-\phi_b)$ and $\sum _{k=0}^q \binom{q}{2k+1}$ elements with  $sin(\phi_a-\phi_b)$. If $q$ is odd (for the given state$|\psi\ran$) the number of cosines and the number of sines are equal. When $q$ is even the number of cosines exceeds by 1. Finally  we get $$|| \mathcal{T}^{(N)}_{|\psi\ran}||^2 = (\pm |a|^2 \pm |b|^2)^2 + 4 |a|^2 |b|^2 cos^2 (\phi_a-\phi_b) \sum_{k=0}^q \binom{q}{2k}+4 |a|^2 |b|^2 sin^2 (\phi_a-\phi_b) \sum_{k=0}^q \binom{q}{2k+1}$$
Note that, using  $|a|^2+|b|^2 =1$, it is easy to see that $||\mathcal{T}^{(N)}_{|\psi\ran}|| \ge 1$, showing that $E_{\mathcal{T}} \ge 0.$
 Next we consider $$|\psi\ran \lan \psi|= |a|^2 \arrowvert b_1\cdots b_N  \ran  \lan b_1\cdots b_N \arrowvert + |b|^2 |b^{\prime}_1 \cdots b^{\prime}_N \ran \lan b^{\prime}_1 \cdots b^{\prime}_N | + a b^* \arrowvert b_1\cdots b_N  \ran \lan b^{\prime}_1 \cdots b^{\prime}_N | $$
 $$+ a^* b |b^{\prime}_1 \cdots b^{\prime}_N \ran  \lan b_1\cdots b_N \arrowvert $$ 
 and trace out the $N$th qubit to get the $N-1$ qubit reduced density matrix  
 $$\rho = |a|^2 \arrowvert b_1\cdots b_{N-1}  \ran  \lan b_1\cdots b_{N-1} \arrowvert + |b|^2 |b^{\prime}_1 \cdots b^{\prime}_{N-1} \ran \lan b^{\prime}_1 \cdots b^{\prime}_{N-1} | $$
 $$+  a b^* \arrowvert b_1\cdots b_{N-1}  \ran \lan b^{\prime}_1 \cdots b^{\prime}_{N-1} | \lan b_N|b^{\prime}_N\ran + a^* b |b^{\prime}_1 \cdots b^{\prime}_{N-1} \ran  \lan b_1\cdots b_{N-1} \arrowvert \lan b^{\prime}_N|b_N\ran $$ 
 Now $$t_{\alpha_1\cdots \alpha_{N-1}}= Tr(\rho \si_{\alpha_1}\otimes \si_{\alpha_2}\otimes \cdots \si_{\alpha_{N-1}})
 = \arrowvert a\arrowvert^2 \lan b_1\cdots b_{N-1}\arrowvert\si_{\alpha_1}\otimes\cdots\otimes\si_{\alpha_{N-1}}\arrowvert b_1\cdots b_{N-1}\ran$$ 
$$ +|b|^2\lan b^{\prime}_1 \cdots b^{\prime}_{N-1}|\si_{\alpha_1}\otimes\cdots\otimes\si_{\alpha_{N-1}}|b^{\prime}_1 \cdots b^{\prime}_{N-1}\ran   +a^* b \lan b_1\cdots b_{N-1}\arrowvert\si_{\alpha_1}\otimes\cdots\otimes\si_{\alpha_{N-1}}|b^{\prime}_1 \cdots b^{\prime}_{N-1}\ran  \lan b_N|b^{\prime}_N\ran  $$
 $$+a b^* \lan b^{\prime}_1 \cdots b^{\prime}_{N-1}|\si_{\alpha_1}\otimes\cdots\otimes\si_{\alpha_{N-1}}\arrowvert b_1\cdots b_{N-1}\ran  \lan b^{\prime}_N|b_N\ran$$
 
 We have for $N-1$ tensor product operators $ \si_3\otimes \si_3\otimes \cdots \otimes \si_3 |b_1 \cdots b_{N-1}\ran=\pm |b_1\cdots b_{N-1}\ran.$ 
 We construct the operators $D$ and $D_k$ corresponding to $B$ and $B_k$ acting on $N-1$ qubits. We then get $D |b_1 \cdots b_{N-1}\ran= \pm |b^{\prime}_1 \cdots b^{\prime}_{N-1}\ran $ and 
  $ D_k |b_1 \cdots b_{N-1}\ran= \pm (i)^k |b^{\prime}_1 \cdots b^{\prime}_{N-1}\ran $ 
 Now, the nonzero elements of $\mathcal{T}^{(N-1)}_{\rho}$ are  
 $t_{33\cdots 3}=\pm |a|^2 \pm |b|^2$,
 $ t_D= \pm a b^* \lan b_N|b^{\prime}_N\ran \pm a^* b \lan b^{\prime}_N | b_N\ran = 2 |a| |b| |\lan b^{\prime}_N | b_N\ran| cos (\phi_a-\phi_b-\alpha)$,   
 
 \begin{displaymath}
 t_{D_k}= \pm (i)^k a b^*  \lan b_N|b^{\prime}_N\ran \pm  (-i)^k a^* b \lan b^{\prime}_N | b_N\ran
 =\left\{ \begin{array}{ll}
 \pm 2 |a| |b|\;|\lan b^{\prime}_N | b_N\ran| cos (\phi_a-\phi_b-\alpha) & \textrm{ if $k$ is even} \\

\pm 2 |a| |b|\; |\lan b^{\prime}_N | b_N\ran| sin (\phi_a-\phi_b-\alpha) & \textrm{ if $k$ is odd} \\
\end{array} \right.
\end{displaymath}
  Finally we get  $$|| \mathcal{T}^{(N-1)}_{\rho}||^2 = (\pm |a|^2 \pm |b|^2)^2 + 4 |a|^2 |b|^2 |\lan b^{\prime}_N | b_N\ran|^2 cos^2 (\phi_a-\phi_b-\alpha) \sum_{k=0}^{q'} \binom{q'}{2k}$$
  $$+4 |a|^2 |b|^2 |\lan b^{\prime}_N | b_N\ran|^2 sin^2 (\phi_a-\phi_b-\alpha) \sum_{k=0}^{q'} \binom{q'}{2k+1}.$$  
  where $q' \le q$ is the number of $\si_1$ operators in $D$. Since $|\lan b^{\prime}_N | b_N\ran|^2 \le 1$ we see that $$|| \mathcal{T}^{(N-1)}_{\rho}||^2 \le || \mathcal{T}^{(N)}_{|\psi\ran}||^2$$  
  equality occurring when $|b_N\ran =|b^{\prime}_N\ran$ in which case $|\psi\ran=|\phi\ran\otimes|b_N\ran.$  
  It is straightforward, but tedious to elevate is proof for the general case 
  $$|\psi\ran=\sum_{\alpha_1 \cdots \alpha_N} a_{\alpha_1 \cdots \alpha_N} |b_{\alpha_1}\cdots b_{\alpha_N}\ran, \; \alpha_i=0,1$$  
  Basically we have to keep track of $\binom{r}{2}$ $B$ type of operators, where $r$ is the number of terms in the expansion of $|\psi\ran$, in order to obtain all nonzero elements of $\mathcal{T}^{(N)}_{|\psi\ran}$. When $N$th particle is traced out, the corresponding elements of $\mathcal{T}^{(N-1)}_{\rho}$ get multiplied by the overlap amplitudes, which leads to the required result.\hfill $\blacksquare$\\

 
 \textit{Continuity of $E_{\mathcal{T}}$}: We show that for $N$-qubit pure states $||(|\psi\ran\lan\psi|-|\phi\ran\lan\phi|)||\rightarrow 0 \Rightarrow \Big{|}E_{\mathcal{T}}(|\psi\ran)-E_{\mathcal{T}}(|\phi\ran)\Big{|}\rightarrow 0 $
 
 \textit{Proof.} $||(|\psi\ran\lan\psi|-|\phi\ran\lan\phi|)||\rightarrow 0 $
$\Rightarrow ||\mathcal{T}^{(N)}_{|\psi\ran}-\mathcal{T}^{(N)}_{|\phi\ran}||\rightarrow 0 $

But  $||\mathcal{T}^{(N)}_{|\psi\ran}-\mathcal{T}^{(N)}_{|\phi\ran}|| \ge \Big{|}||\mathcal{T}^{(N)}_{|\psi\ran}||-||\mathcal{T}^{(N)}_{|\phi\ran}||\Big{|}$

Therefore $||\mathcal{T}^{(N)}_{|\psi\ran}-\mathcal{T}^{(N)}_{|\phi\ran}||\rightarrow 0 \Rightarrow \big{|}||\mathcal{T}^{(N)}_{|\psi\ran}||-||\mathcal{T}^{(N)}_{|\phi\ran}||\big{|} \rightarrow 0$
  
   $\Rightarrow \Big{|}E_{\mathcal{T}}(|\psi\ran)-E_{\mathcal{T}}(|\phi\ran)\Big{|}\rightarrow 0. $ \hfill $\blacksquare$\\

 \bc  
\textbf{A. Entanglement of multiple copies of a given state}\\
 \ec

{\it LU invariance. } We show that  $E_{\mathcal{T}}$ for multiple copies of $N$-qubit pure state $|\psi\ran$ is $LU$ invariant.
Consider a system of $N\times k$ qubits in the state $|\chi\ran=|\psi\ran\otimes |\psi\ran \otimes \cdots\otimes |\psi\ran$ ($k$ copies). It is straightforward to check that [24]

$$  \mathcal{T}^{(N)}_{|\chi\ran}= \mathcal{T}^{(N)}_{|\psi\ran}\circ \mathcal{T}^{(N)}_{|\psi\ran}\circ \cdots \circ \mathcal{T}^{(N)}_{|\psi\ran} \eqno{(29)}$$
   
 This implies, in a straightforward way, that $$|| \mathcal{T}^{(N)}_{|\chi\ran}||=||  \mathcal{T}^{(N)}_{|\psi\ran}||^k.$$
    Since by proposition 6 $||  \mathcal{T}^{(N)}_{|\psi\ran}||$ is $LU$ invariant, so is $|| \mathcal{T}^{(N)}_{|\chi\ran}||$.
    
    Let $|\psi\ran$ be a $N$-qubit pure state and $|\chi\ran=|\psi\ran\otimes|\psi\ran$. Then $E_{\mathcal{T}}(|\chi\ran)$ is expected to satisfy $$E_{\mathcal{T}}(|\chi\ran)\ge E_{\mathcal{T}}(|\psi\ran)$$
    
    We again use the fact that $$  \mathcal{T}^{(N)}_{|\chi\ran}= \mathcal{T}^{(N)}_{|\psi\ran}\circ \mathcal{T}^{(N)}_{|\psi\ran}$$
    
    which gives $$|| \mathcal{T}^{(N)}_{|\chi\ran}||=||  \mathcal{T}^{(N)}_{|\psi\ran}||^2$$
    
    Since $||  \mathcal{T}^{(N)}_{|\psi\ran}||\ge 1$ we get    
   $ || \mathcal{T}^{(N)}_{|\chi\ran}||\ge ||  \mathcal{T}^{(N)}_{|\psi\ran}||$ or,  $$E_{\mathcal{T}}(|\chi\ran)\ge E_{\mathcal{T}}(|\psi\ran).$$

 \textit{Superadditivity }: We have to show, for $N$qubit states   $|\psi\ran$ and $|\phi\ran$ that $$E_{\mathcal{T}}(|\psi\ran \otimes |\phi\ran )\ge E_{\mathcal{T}}(|\psi\ran)+ E_{\mathcal{T}}(|\phi\ran). \eqno{(30)}$$  
We already know that for $|\chi\ran=|\psi\ran\otimes|\phi\ran$
$$|| \mathcal{T}^{(N)}_{|\chi\ran}||= ||  \mathcal{T}^{(N)}_{|\psi\ran}||\; ||  \mathcal{T}^{(N)}_{|\phi\ran}||$$

Thus Eq. (30) gets transformed to $$||  \mathcal{T}^{(N)}_{|\psi\ran}||\; ||  \mathcal{T}^{(N)}_{|\phi\ran}||-1 \ge ||  \mathcal{T}^{(N)}_{|\psi\ran}||+ ||  \mathcal{T}^{(N)}_{|\phi\ran}||-2$$

which is true for $||  \mathcal{T}^{(N)}_{|\psi\ran}||\ge 1$ and $||\mathcal{T}^{(N)}_{|\phi\ran}||\ge 1$. \hfill $\blacksquare$\\

\bc
\textbf{B. Computational considerations}
\ec

Computation or experimental determination of $E_{\mathcal{T}}$ involves $3^{N}$ elements of $\mathcal{T}^{(N)}$ so that it increases exponentially with the number of qubits $N$. However, for many important classes of states, $E_{\mathcal{T}}$ can be easily computed and increases only polynomially with $N$. We have already computed $E_{\mathcal{T}}$ for the class of $N$ qubit $W$ states, $GHZ$ states and their superpositions. We have also computed  
$E_{\mathcal{T}}$ for an important physical system like $1D$ Heisenberg antiferromagnet. For symmetric or antisymmetric states $\mathcal{T}^{(N)}$ is supersymmetric, that is, the value of its elements are invariant under any permutation of its indices [24]. This reduces the problem to the computation of $\fr{1}{2}(N+1)(N+2)$ distinct elements of $\mathcal{T}^{(N)},$ which is quadratic in $N$.\\

\bc

\textbf{C. Entanglement dynamics : Grover algorithm}

\ec
We show that $E_{\mathcal{T}}$ can quantify the evolution of entanglement. We consider Grover's algorithm. The goal of Grover's algorithm is to convert the initial state of $N$ qubits, say $|0\cdots 0\ran,$ to a state that has probability bounded above $1/2$ of being in the state $|a_1\cdots a_{N}\ran,$ using       
$$U_a|b_1\cdots b_{N}\ran=(-1)^{\Pi\delta_{a_jb_j}}|b_1\cdots b_{N}\ran$$
fewest times possible. Grover showed that this can be done with $O(\sq{2^{N}})$ uses of $U_a$ by starting with the state $$\fr{1}{\sq{2^{N}}}\sum_{x=0}^{2^{N}-1}|x\ran \;=\;H^{\otimes N}|0\cdots0\ran$$ where 

\begin{displaymath}
H=\fr{1}{\sq{2}}
\left(\begin{array}{cc}
1& 1  \\
1 & -1\\
\end{array}\right)
\end{displaymath}
and then iterating the transformation $H^{\otimes N}U_a H^{\otimes N}U_a$ on this state [23]. The initial state is a product state as is the target state, but intermediate states $\psi(k)$ are entangled for $k>0$ iterations. Figure 6 shows the development of $E_{\mathcal{T}}(|\psi(k)\ran)$ with number of iterations $k,$ for six qubits. The values of $k$ for which $E_{\mathcal{T}}$ vanishes are the iterations at which the probability of measuring $|a_1\cdots a_{N}\ran$ is close to $1.$ Thus $E_{\mathcal{T}}$ can be used to quantify the evolution of a $N$-qubit entangled state.\\

\begin{figure}[!ht]
\begin{center}
\includegraphics[width=12cm,height=8cm]{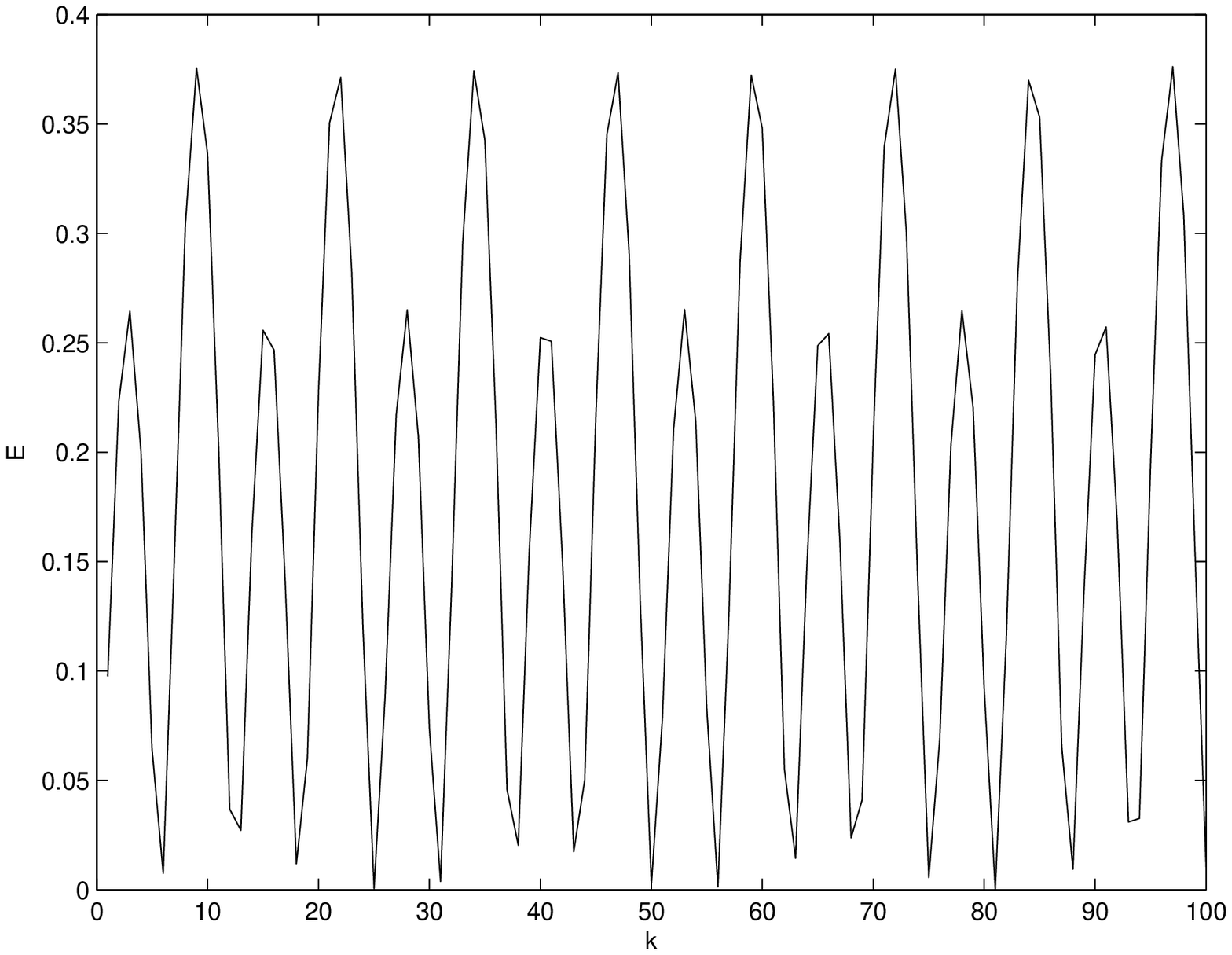}

figure 6. Entanglement in Grover's algorithm for six qubits as a function of number of iterations. 
\end{center}
\end{figure}

\bc
\textbf{V. EXTENSION TO MIXED STATES}\\

\ec

The extension of $E_{\mathcal{T}}$ to mixed states $\rho$ can be made via the use of the {\it convex roof} or {\it (hull)} construction as was done for the entanglement of formation [16]. We define $E_{\mathcal{T}}(\rho)$ as a minimum over all decompositions $\rho=\sum_{i}p_i\arrowvert\psi_i\ran\lan\psi_i\arrowvert$ into pure states i.e.
$$E_{\mathcal{T}}(\rho)=\min_{\begin{subarray}{I} 
  \hskip  .1cm  {\{p_i,\psi_i\}}
 \end{subarray}}
\sum_{i}p_iE_{\mathcal{T}}(\arrowvert\psi_i\ran).\eqno{(31)}$$

The existence and uniqueness of the convex roof for $E_{\mathcal{T}}$ is guaranteed because it is a continuous function on the set of pure states [39]. This entanglement measure is expected to satisfy conditions (a), (b) and (c) given in section IV and is expected to be  
(d)convex under discarding of information, i.e. $$\sum_{i}p_i E_{\mathcal{T}}(\rho_i)\geq E_{\mathcal{T}}(\sum_{i}p_i\rho_i). \eqno{(32)}$$ 

The criteria (a)-(d) above are considered to be the minimal set of requirements for any entanglement measure so that it is an entanglement monotone [29].

Evidently, criteria (a) and (b) are satisfied by $E_{\mathcal{T}}(\rho)$ defined via convex roof as it is satisfied by
$E_{\mathcal{T}}$ for pure states. Condition (d) follows from the fact that every convex hull (roof) is a convex function [40]. We need to prove (c), which is summarized in the following proposition.
 
\textit{Proposition 9}: If  a $N$-qubit mixed state $\rho$ is subjected to a local operation on $i$ th qubit giving outcomes $k$ with probabilities $p_k$ and leaving residual $N$ qubit mixed state $\rho_k$, then the expected entanglement $\sum_{k} p_k E_{\mathcal{T}}(\rho_k)$ of the residual state is not greater than the entanglement $E_{\mathcal{T}}(\rho)$ of the original state. $$\sum_{k} p_k E_{\mathcal{T}}(\rho_k) \le E_{\mathcal{T}}(\rho)$$ (If the operation is simply throwing away part of the system, then there will be only one value of $k$, with unit probability.)

The proof follows from the monotonicity of $E_{\mathcal{T}}(|\psi\ran)$  for pure states that is propositions 6, 7 and 8.
Bennett et al. prove a version of proposition 9 in [34], which applies to any measure satisfying propositions 6,7 and 8. Thus the same proof applies to proposition 9, so we skip it. 

Note that any sequence of local operations comprises local operations drawn from the set of basic operations (i)-(iv) above, so that proposition 9 applies to any such sequence. Thus we can say that expected entanglement of a $N$-qubit system, measured by $E_{\mathcal{T}}(\rho)$, does not increase under local operations. \\

\bc
\textbf{VI. A RELATED ENTANGLEMENT MEASURE}
\ec

We consider following entanglement measure. Consider $$\mathcal{E}_{\mathcal{T}}(|\psi\ran)=log_2||\mathcal{T}^{(N)}||=log_2(E_{\mathcal{T}}(|\psi\ran)+1)$$

where $\mathcal{T}^{(N)}$ is the $N$-way correlation tensor occuring in the Bloch representation of $\rho=|\psi\ran\lan\psi|.$ 

Proofs of propositions 1 through 8 easily go through for $\mathcal{E}_{\mathcal{T}}(|\psi\ran)$. We prove continuity as follows.

\textit{Continuity of $\mathcal{E}_{\mathcal{T}}(|\psi\ran)$} : We have to show, for two $N$-qubit states $|\psi\ran$ and $|\phi\ran$ that $||(|\psi\ran\lan\psi|-|\phi\ran\lan\phi|)||\rightarrow 0 \Rightarrow \Big{|}\mathcal{E}_{\mathcal{T}}(|\psi\ran)-\mathcal{E}_{\mathcal{T}}(|\phi\ran)\Big{|}\rightarrow 0 $
 
we have,  $||(|\psi\ran\lan\psi|-|\phi\ran\lan\phi|)||\rightarrow 0 $
$\Rightarrow ||\mathcal{T}^{(N)}_{|\psi\ran}-\mathcal{T}^{(N)}_{|\phi\ran}||\rightarrow 0 $

But  $ ||\mathcal{T}^{(N)}_{|\psi\ran}-\mathcal{T}^{(N)}_{|\phi\ran}|| \ge \big{|}||\mathcal{T}^{(N)}_{|\psi\ran}||- ||\mathcal{T}^{(N)}_{|\phi\ran}||\big{|}$

Further, whenever $|||\mathcal{T}^{(N)}_{|\psi\ran}|| \ge 1$ and $|\mathcal{T}^{(N)}_{|\phi\ran}|| \ge 1$

we have $\big{|}||\mathcal{T}^{(N)}_{|\psi\ran}||- ||\mathcal{T}^{(N)}_{|\phi\ran}||\big{|} \ge \big{|}log_2(||\mathcal{T}^{(N)}_{|\psi\ran}||)- log_2(||\mathcal{T}^{(N)}_{|\phi\ran}||)\big{|}$

Thus $ ||\mathcal{T}^{(N)}_{|\psi\ran}-\mathcal{T}^{(N)}_{|\phi\ran}||\rightarrow 0 \Rightarrow \big{|}||\mathcal{T}^{(N)}_{|\psi\ran}||- ||\mathcal{T}^{(N)}_{|\phi\ran}||\big{|} \rightarrow 0 \Rightarrow \big{|}log_2(||\mathcal{T}^{(N)}_{|\psi\ran}||)- log_2(||\mathcal{T}^{(N)}_{|\phi\ran}||)\big{|} \rightarrow 0 \Rightarrow \Big{|}\mathcal{E}_{\mathcal{T}}(|\psi\ran)-\mathcal{E}_{\mathcal{T}}(|\phi\ran)\Big{|}\rightarrow 0 .$

However, $\mathcal{E}_{\mathcal{T}}(|\psi\ran)$ has the added advantage that it is additive ( while $E_{\mathcal{T}}(|\psi\ran)$ is superadditive).
Indeed, from section IV A we see that for $k$ copies
$$\mathcal{E}_{\mathcal{T}}(|\psi\ran\otimes|\psi\ran\otimes\cdots \otimes |\psi\ran)=k \mathcal{E}_{\mathcal{T}}(|\psi\ran)$$

Similarly $\mathcal{E}_{\mathcal{T}}(|\psi\ran\otimes|\phi\ran)=\mathcal{E}_{\mathcal{T}}(|\psi\ran)+\mathcal{E}_{\mathcal{T}}(|\phi\ran)$

The extension of $\mathcal{E}_{\mathcal{T}}(|\psi\ran)$ to mixed states via convex roof construction is similar to that of $E_{\mathcal{T}}(|\psi\ran).$ Thus $\mathcal{E}_{\mathcal{T}}(|\psi\ran)$ has all the properties of $E_{\mathcal{T}}(|\psi\ran)$, with an additional property that $\mathcal{E}_{\mathcal{T}}(|\psi\ran)$ is additive, while $E_{\mathcal{T}}(|\psi\ran)$ is superadditive.\\

\bc

\textbf{VII. CONCLUSION}
\ec
In conclusion, we have developed an experimentally viable entanglement measure for $N$-qubit pure states, which passes almost all the tests for being a good entanglement measure. This is a global entanglement measure in the sense that it does not involve partitions or cuts of the system in its definition or calculation. This measure has quadratic computational complexity for symmetric or antisymmetric states. Computational tractability is not a serious problem if $N$ is not too large and the measure can be easily computed for systems comprising small number of qubits, which can have many important applications such as teleportation of multiqubit states, quantum cryptography, dense coding, distributed evaluation of functions [41] etc. However, finding other classes of states for which $E_{\mathcal{T}}$ can be computed polynomially will be useful. It will be very interesting to seek applications of this measure to situations 
like quantum phase transitions [11], transfer of entanglement along spin chains [42], NOON states in quantum lithography [43] etc.
Finally, we have extended our measure to the mixed states and established its various properties, in particular, its monotonicity. We may also note that neither its definition, nor its properties depend, in an essential way, on the fact that we are dealing with qubits, so that this measure can be defined and applied to a general $N$-partite quantum system.   \\

\bc
\textbf{ACKNOWLEDGMENTS}
\ec
We thank Guruprasad Kar and Professor R. Simon for encouragement. It is a pleasure to thank Professor Sandu Popescu for an illuminating discussion and encouragement. We thank Sougato Bose for his helpful suggestion.  We thank Sibasish Ghosh and Pranaw Rungta for thier critical comments. ASMH thanks Sana'a University for financial support.

\bc

\textbf{References}\\

\ec
\begin{verse} 

[1] C. H. Bennett, G. Brassard, C. Cr\'epeau, R. Jozsa, A. Peres,
and W. K. Wootters, Phys. Rev. Lett. \textbf{70}, 1895 (1993).

[2] Ye Yeo and Wee Kang Chua, Phys. Rev. Lett. \textbf{96}, 060502 (2006).

[3] C. H. Bennett and G. Brassard (unpublished); D. Deutsch, A. Ekert, R. Jozsa, C. Macchiavello, S. Popescu, and A. Sanpera, Phys. Rev. Lett. \textbf{77} 2818, (1996), \textbf{80}, 2022 (1998), H.-K. Lo, {\it in Introduction to Quantum Computation and Information}, edited by H.-K. Lo, S. Popescu and T. Spiller (World Scientific,
Singapore, 1998), pp. 76–119; H. Zbinden {\it ibid.} pp. 120–142.

[4] C. H. Bennett and S. J. Wiesner, Phys. Rev. Lett. 69, 2881 (1992).

[5] C. H. Bennett, C. A. Fuchs, and J. A. Smolin, in {\it Quantum
Communication, Computing and Measurement}, edited by O. Hirota, A. S. Holevo, and C. M. Caves (Plenum, New York, 1997).

[6] C. H. Bennett, P. W. Shor, J. A. Smolin, and A. V. Thapliyal,
Phys. Rev. Lett. \textbf{83}, 3081 (1999).

[7] P. W. Shor, Phys. Rev. A \textbf{52}, R2493 (1995); D. Gottesman,
Ph.D. Thesis, California Institute of Technology, 1997; LANL
e-print, quant-ph/9705052.

[8] D. Deutsch, Proc. R. Soc. London, Ser. A \textbf{400}, 97 (1985); D.
Deutsch ibid., \textbf{425}, 73 (1989); M. A. Nielsen and I. L. Chuang, {\it Quantum Computation and Quantum Information}, (Cambridge University Press 2000).

[9] L. K. Grover, LANL e-print, quant-ph/9704012.

[10] R. Cleve and H. Buhrman, LANL e-print quant-ph/9704026.

[11] T. J. Osborne and M. A. Nielsen, Phys. Rev. A  \textbf{66}, 032110 (2002); A. Osterloh et al., Nature (London)  \textbf{416}, 608 (2002).

[12] M. B. Plenio, S. Virmani, Phys. Rev. Lett. \textbf{99}, 120504 (2007)

[13] M. B. Plenio and S. Virmani , Quantum Inf. Comput., Vol.  \textbf{7}, 1 (2007).

[14] K. \.Zyczkowski and I. Bengstsson, quant-ph/0606228.

[15] R., P., M., K. Horodecki, quant-ph/0702225v2.

[16] W.K. Wootters, Phys. Rev. Lett. \textbf{80}, 2245 (1998).

[17] K. G. H. Vollbrecht and R.F. Werner, Phys. Rev. A \textbf{64}, 062307 (2001).

[18] B.M. Terhal and K.G.H. Vollbrecht, Phys. Rev. Lett. \textbf{85}, 2625 (2000).

[19] A. R. Usha Devi, R. Prabhu, and A. K. Rajagopal, Phys. Rev. Lett.   \textbf{98}, 060501 (2007).\\

[20] V. Vedral and M.B. Plenio, Phys. Rev. A \textbf{57}, 1619 (1998).

[21] K. \.Zyczkowski, P. Horodecki, A. Sanpera, and M. Lewenstein,
Phys. Rev. A \textbf{58}, 883 (1998); G. Vidal and R.F. Werner, ibid.
\textbf{65}, 032314 (2002).

[22]  J. Eisert and H.J. Briegel, Phys. Rev. A \textbf{64}, 022306 (2001).

[23] David A. Meyer and Nolan R. Wallach, J. Math. Phys.  \textbf{43}, 4273 (2002).

[24] Ali Saif M. Hassan and Pramod S. Joag, quant-ph/0704.3942, to appear in QIC.

[25] G. Kimura and A. Kossakowski, Open Sys. Inf. Dyn., Vol. \textbf{12}, 207 (2005).

[26] G. Kimura, Phys. Lett. A  \textbf{314}, 339 (2003).

[27] M.S. Byrd and N. Khaneja, Phys. Rev. A  \textbf{68}, 062322 (2003).

[28] Tzu-Chieh Wei and Paul M. Goldbart, Phys. Rev. A   \textbf{68}, 042307 (2003).

[29] G. Vidal, J. Mod. Opt.  \textbf{47}, 355 (2000).

[30] I. Bengtsson and K. \.Zyczkowski, {\it Geometry of Quantum States}, (Cambridge University Press 2006).

[31] T. G. Kolda, {\it Multilinear operators for higher-order decompositions}, Tech. Report SAND2006-2081, Sandia National Laboratories, Albuquerque, New Mexico and
Livermore, California, Apr. 2006.

[32] T. G. Kolda,  SIAM J. Matrix Anal. A., Vol. \textbf{23}, 243 (2001).

[33] A. Acin, A. Andrianov, L. Costa, E. Jane, J. I. Latorre and R. Tarrach, Phys. Rev. Lett. \textbf{85}, 1560 (2000).

[34] C. Bennett, D. P. DiVincenzo, J. A. Smolin, and W. K. Wootters, Phys. Rev.  A \textbf{54}, 3824 (1996).

[35] L. De Lathauwer, B. De Moor, and J. Vandewalle, SIAM J. Matrix Anal. A., Vol. \textbf{21}, 1253 (2000).

[36] Jing Zhang, Chun-Wen Li, Jian-Wu Wu, Re-Bing Wu and Tzyh-Jong Tam, Phys. Rev. A, \textbf{73}, 022319 (2006).

[37] R.A. Horn and C.R. Johnson, {\it  Matrix Analysis}, Cambridge University Press (Cambridge 1985).

[38] Yan-Kui Bai, Dong and Z. D. Wang, Phys. Rev. A \textbf{76}, 022336 (2007).

[39] A. Uhlmann, Phys. Rev. A \textbf{62}, 032307 (2000).

[40] A. Uhlmann, LANL e-print, quant-ph/9704017v2.

[41] G. Brassard, quant-ph/0101005.

[42] A. Bayat, S. Bose, quant-ph/0706.4176.

[43] A. N. Boto, P. Kok, D. S. Abrame, S. L. Braunstein, C. P. Williams and J. P. Dowling, Phys. Rev. Lett. \textbf{85}, 2733 (2000).

\end{verse}

\end{document}